\documentclass[11pt,aps,onecolumn,preprintnumbers,floatfix,superscriptaddress]{revtex4-2}

\usepackage[final]{graphicx}
\usepackage{hyperref}
\usepackage{amsmath}
\usepackage{bbm}
\usepackage{amsfonts}
\usepackage{amssymb}
\usepackage{latexsym}
\usepackage{graphicx}
\usepackage[english]{babel}
\usepackage{multirow}
\usepackage{float}
\usepackage{url}
\usepackage{slashed}
\usepackage{xcolor}
\usepackage[utf8]{inputenc}
\usepackage{verbatim}
\usepackage{lipsum}

\newcommand{\rmi}[1]{{\mbox{\scriptsize #1}}}
\newcommand{\rmii}[1]{{\mbox{\tiny\rm{#1}}}}
\newcommand{\mD}{m_\rmii{D}}
\newcommand{\mh}{\mu_{h}}
\newcommand{\bmu}{\bar\mu}
\newcommand{\g}{g}
\newcommand{\gp}{g'}
\newcommand{\gY}{g_{Y}}
\newcommand{\gs}{g_\rmi{s}}

\newcommand{\Nf}{N_{\rm f}}

\newcommand{\gammaE}{{\gamma_\rmii{E}}}
\newcommand{\beq}{\begin{equation}}
\newcommand{\eeq}{\end{equation}}
\newcommand{\bi}{\begin{itemize}}
\newcommand{\ei}{\end{itemize}}

\newcommand{\mc}[1]{\mathcal{#1}}

\newcommand{\be}{\begin{equation}}
\newcommand{\ee}{\end{equation}}
\newcommand{\ba}{\begin{array}}
\newcommand{\ea}{\end{array}}
\newcommand{\bea}{\begin{eqnarray}}
\newcommand{\eea}{\end{eqnarray}}

\newcommand{\nn}{\nonumber}

\makeatletter
\renewcommand{\maketag@@@}[1]{\hbox{\m@th\normalsize\normalfont#1}}%
\makeatother

\newlength{\halfpagewidth}
\divide\halfpagewidth by 2

\begin{document}
\title{First-order Electroweak phase transition at finite density}

\author{Renhui Qin}
\email{20222701021@stu.cqu.edu.cn}
\affiliation{Department of Physics and Chongqing Key Laboratory for Strongly Coupled Physics, Chongqing University, Chongqing 401331, P. R. China}

\author{Ligong Bian\footnote{Corresponding Author.}}
\email{lgbycl@cqu.edu.cn}
\affiliation{Department of Physics and Chongqing Key Laboratory for Strongly Coupled Physics, Chongqing University, Chongqing 401331, P. R. China}
\affiliation{ Center for High Energy Physics, Peking University, Beijing 100871, China}

\begin{abstract}
We study the Electroweak phase transition with the Standard Model effective field theory at finite temperature and finite density. Utilizing the dimensional reduction approach, we construct the tree dimensional thermal effective field theory at finite density and investigate the phase transition dynamics. We evaluate how the results depend on the renormalization scale and the chemical potential.
 Our results show that, with the tree dimensional thermal effective potential at 2-loop level, we can effectively reduce the theoretical uncertainty in the calculations of the phase transition parameters due to the renormalization scale dependence,
 and the new physics scale is restricted to be
 $\Lambda\lesssim (770-800)$ GeV by the
 baryon number washout avoidance condition.
 Meanwhile, the presence of the chemical potential would affect the phase transition parameter and make the constraints from the baryon number washout avoidance condition more strict, especially for weaker first-order phase transition scenarios at higher new physics scales.

\end{abstract}

\maketitle
\tableofcontents
\date{\today}
\newpage

\section{Introduction}
First-order phase transition (PT) is a common prediction of many particle physics models beyond the Standard Model~\cite{Caldwell:2022qsj,Athron:2023xlk,Bian:2021ini}, such transitions are a crucial ingredient for an explanation of the baryon asymmetry of the Universe through {\it electroweak baryogenesis} (EWBG)\cite{Kuzmin:1985mm,Cohen:1993nk,Riotto:1999yt,Morrissey:2012db}, and can produce gravitational wave (GW) signatures to be probed by space-based interferometers\cite{Kamionkowski:1993fg,Apreda:2001us,Alves:2018jsw}.
When new physics is integrated out, one arrives at the standard model effective field theory (SMEFT), where
the potential barrier between symmetric and broken phase can appear at tree-level potential, which is mostly captured by the dimension-six operator of $(\Phi^\dag\Phi)^3/\Lambda^2$ with
the $\Lambda$ indicating the new physics (NP) scale. Previously, there are many studies on the electroweak PT in the framework of SMEFT based on four-dimensional finite temperature thermal field theory~\cite{Cai:2017tmh,Croon:2020cgk,Grojean:2004xa,Delaunay:2007wb,Chala:2018ari,Bodeker:2004ws,Hashino:2022ghd,Damgaard:2015con,Postma:2020toi}. These works show that a lot of parameter spaces of SMEFT can predict strong first-order PT to be probed at near future GW detectors.

To analyze the electroweak PT dynamics with four-dimensional finite temperature thermal field theory,
daisy resummation is usually applied to improve the power like infrared divergence (IR) problem~\cite{Farakos:1994xh,Kajantie:1995dw,Farakos:1994kx}.
Another way to study the property of PT, the 3d dimensional reduction (DR) method has been developed, which applies to the
non-perturbative lattice simulation and suffer
less theoretical uncertainty from the
IR problem~\cite{Farakos:1994kx,Farakos:1994xh,Kajantie:1995dw,Gynther:2003za}.
Previous studies show that the DR suffers the uncertainty of the renormalization scale dependence which is reflected in the convergence of perturbation theory~\cite{Schicho:2022wty,Niemi:2021qvp,Croon:2020cgk,Kainulainen:2019kyp}. Meanwhile, the EWBG and detectable GW in the mHz to Hz range require a strong first-order electroweak PT caused by a strong coupling which will reduce the convergence, and its correction is more pronounced at 2-loop level and higher orders. Therefore, in the procedure of calculating electroweak PT with the 3d DR method, at least 2-loop level is needed to reduce the theoretical uncertainty from the renormalization scale dependence.

In this paper, we adopt the DR approach to study the electroweak PT within the framework of the SMEFT at finite density.
To ensure a positive-definite value for the number of bosons, the chemical potential needs less than the mass $\mu_i\leq m_i$\cite{Haber:1981fg}. The DR approach usually needs to integrate heavy mode to build a light theory $(g^2 T)$, so the chemical potential in this work requires less than light scale $\mu_i\leq g^2 T$.
In this work, we consider theoretical uncertainties caused by renormalization scale dependence and the effect of the chemical potential on the dynamics of strong first-order electroweak PT relevant for the GW prediction and EWBG. Since there is only a single parameter (the NP scale $\Lambda$) to characterize the phase transition behavior and the electroweak sphaleron energy, we can easily consider the effect of both the energy scale and chemical potential simultaneously.

The paper is organized as follows. In Section.\ref{drsmeft}, we give a brief introduction to the procedure of the DR approach and construct the thermal SMEFT at finite density, where we elaborate on the definition of the PT parameter and the electroweak sphaleron. In Section.\ref{phasetransition}, we study the dependence of the renormalization scale and the effect of chemical potential on the PT in detail, including the PT parameters relevant for GW prediction and the baryon number washout avoidance condition. In the end, we conclude with the Section.\ref{endsection}. Appendix \ref{Parameter} presents the Beta functions and the input parameters of our calculation. Appendix \ref{resdr} gives the final result of the parameters of the DR approach. Appendix \ref{respotential} gives the effective potential at 2-loop level. Appendix \ref{resdr} shows the effective potential at different $\Lambda$ with critical temperature and nucleation temperature. Appendix~\ref{resspe} gives the detail of our calculation for the electroweak sphaleron energy.

\section{Thermal Standard model effective field theory at finite density}\label{drsmeft}
In this section, we construct the thermal SMEFT at finite density to study the PT dynamics around electroweak scales.
\subsection{The Model}
We use the SMEFT with only a single dimension-six operator to study the first-order PT property. The Euclidean action is defined by
\begin{equation}
S=\int_0^\beta \mathrm{d}\,\tau \int\mathcal{L}\mathrm{d}^3 x\;,
\end{equation}
with
\begin{equation}\label{lagrangian}
\begin{aligned}
\mathcal{L}=&-\frac{1}{4}G_{\mu\nu}^a G_{\mu\nu}^a-\frac{1}{4}F_{\mu\nu}F^{\mu\nu}+(D_\mu\Phi)^\dag(D_\mu\Phi)-V(\Phi)+\overline{q}_L\slashed{D}q_L\\
&+\overline{l}_L\slashed{D}l_L+\overline{e}_R\slashed{D}e_R+\overline{u}_R\slashed{D}u_R+\overline{d}_R\slashed{D}d_R
+g_Y(\overline{q}_L\tilde{\Phi}t_R+\overline{t}_Ru\Tilde{\Phi}^\dag q_L)\;,
\end{aligned}
\end{equation}
and
\begin{equation}
V(\Phi)=\mu_h^2\Phi^\dag \Phi+\lambda (\Phi^\dag \Phi)^2+c_6 (\Phi^\dag \Phi)^3,\quad c_6=\Lambda^{-2}\;.
\end{equation}
Here, the $\Lambda$ is the cut-off scale describing where the NP enters. The covariant derivative is defined as $D_\mu=\partial_\mu-igIA_\mu^i\tau^i-ig^\prime Y B_\mu$, where $I$ and $Y$ are the weak isospin and weak hypercharge of the corresponding doublet/singlet. The $l_L$ and $q_L$ denote the left-hand lepton and quark doublet, and $e_R,u_R,d_R$ denote the right-hand lepton, up-type and down-type quarks. We consider the top quark has non-zero mass with Yukawa coupling $g_Y$, and $\tilde{\Phi}=i\sigma_2 \Phi^*$, where $\Phi$ is the SM Higgs doublet.

To keep the charge conservation in this system, one can choose the hypercharge and the third component of the isospin. The corresponding currents are given as~\cite{Gynther:2003za,Kapusta:1990qc}:
\begin{equation}
\begin{aligned}
j_\mu^Y=&\frac{1}{2}\sum \left(\frac{1}{3}\overline{q}_L\gamma_\mu q_L+\frac{4}{3}\overline{u}_R\gamma_\mu u_R-\frac{2}{3}\overline{d}_R\gamma_\mu d_R-\overline{l}_L\gamma_\mu l_L-2\overline{e}_R\gamma_\mu e_R\right)-\frac{i}{2}[(D_\mu \Phi)^\dag \Phi-\Phi^\dag D_\mu \Phi],\\
j_\mu^3=&\frac{1}{2}\sum(\overline{q}_L\gamma_\mu\tau^3 q_L+\overline{l}_L\gamma_\mu\tau^3 l_L)-\frac{i}{2}[(D_\mu\Phi)^\dag\tau^3\Phi-\Phi^\dag\tau^3D_\mu\Phi]-\epsilon^{3bc}A_\nu^b F_{\mu\nu}^c\;.
\end{aligned}
\end{equation}
Then, the electromagnetic and weak-neutral  currents are
\begin{equation}\label{currents}
\begin{aligned}
N_1=&\int \mathrm{d}^3 x(j_0^Y+j_0^3)\;,\\
N_3=&\int \mathrm{d}^3 x(\sin^2\theta j_0^Y-\cos^2\theta j_0^3)\frac{2}{\cos 2\theta}\;,
\end{aligned}
\end{equation}
where the $\theta$ is the Weinberg angle. In addition, the lepton number currents and baryon number currents are not conserved due to the triangle anomaly\cite{Rubakov:1996vz}. But one can make the $N_f$ conserved linear combinations of the currents for baryon and lepton by define\cite{Gynther:2003za}
\begin{equation}
N_{2,i}=\frac{1}{N_f}B-L_i,\quad i=1,...,N_f
\end{equation}
with
\begin{equation}
\begin{aligned}
B&=\frac{1}{3}\sum_{f,c}\int \mathrm{d}^3 x \overline{q}_{f,c}\gamma_0 q_{f,c}\;,\\
L_i&=\int\mathrm{d}^3 x\left(\overline{e}_i\gamma_0 e_i+\frac{1}{2}\overline{\nu}_i\gamma_0(1-\gamma_5)\nu_i\right).
\end{aligned}
\end{equation}
The $f$ and $c$ are flavor and color for quark fields. Then, the partition function with chemical potential is given by
\begin{equation}
\begin{aligned}
\mathcal{Z}=& \mathrm{Tr} e^{-\beta(H-\mu_k N_k)}\\
=&\mathrm{Tr} e^{-\beta(H-\mu_{i}N_{2,i}-\mu_{Y}^\prime N_1-\mu_{T_3} N_3)}\\
=&\int \mathcal{D}\varphi e^{-S+\int_0^\beta\mathrm{d}\tau \sum_{i}^{N_f}\mu_iN_{2,i}}\;,
\end{aligned}
\end{equation}
with
\begin{equation}
\sum_{i}^{N_f}\mu_iN_{2,i}=\frac{1}{N_f}(\mu_1+\mu_2+\mu_3)B-(\mu_1 L_1+\mu_2 L_2+\mu_3 L_3)\;.
\end{equation}
Here, the $S$ contains the gauge fixing and ghost terms. After redefining  the temporal components of the gauge fields
\begin{equation}
    g^\prime B_0\rightarrow g^\prime B_0-i \mu_Y,\quad g A_0^3\rightarrow g A_0^3-i \mu_{T^3},
\end{equation}
the action given above becomes the standard electroweak action and $\mu_Y,\mu_{T^3}$ are not explicit in the path integral anymore, which means that the chemical potentials and the static
modes of the temporal components of
gauge fields enter the path integral in the same way.

Assuming that all the chemical potentials of the leptons have the same value
\begin{equation}\label{BLasum}
\mu_B=\frac{1}{N_f}\sum_{i=1}^{N_f}\mu_i,\quad \mu_{L_i}=-\mu_i\;,
\end{equation}
which results in $N_f\mu_B+\sum_i\mu_i=0$ which can remove the Chern-Simons terms that appear
due to B+L nonconservation.
We then have
\begin{equation}\label{partitionfun}
\mathcal{Z}=\int \mathcal{D}\varphi e^{-S+\int_0^\beta\mathrm{d}\tau\left(\mu_B B+\sum_{i=1}^{n_f}\mu_{L_i}L_i\right)}\;.
\end{equation}

\subsection{Dimensional Reduction}

The path integral given by Eq.\eqref{partitionfun} can not be evaluated with perturbation theory because of the IR divergence caused by bosonic zero modes. The other modes are massive with $m\sim \pi T$, so they can be integrated out safely. For this reason, we use the DR approach to include the contribution of the bosonic zero modes.
DR can be described as equilibrium thermodynamics of a weakly interacting quantum field theory in the Matsubara formalism of
\begin{equation}\label{matsubara}
\phi(x)=T\sum_{n=-\infty }^{\infty}\int \frac{\mathrm{d}^3\, p}{(2\pi)^3}\phi(\omega_n,\mathbf{p})e^{i\omega_n \tau}e^{-i\mathbf{p\cdot x}}
\end{equation}
where $\omega_n=2 n \pi  T$ for bosons and $\omega_n=(2n+1)\pi T$ for fermions.  This theory works when the coupling constant is small $g<1$, and the couplings are dimensionful. The structure of perturbative expansion is modified by resummations. To build the theory, one needs a matching between 3d and 4d fields and parameters.
The matching in DR needs to integrate out heavier Matsubara mode. Thus we have to define the energy scales in this theory. There are three different energy scales $\pi T$,$g T^2$, and $g^2 T^2$, named superheavy, heavy, and light mode (or heavy, soft, and ultrasoft mode, but we don't use it in this paper). The superheavy modes are fermions and non-zero boson Matsubara modes with mass $\sim\pi T$, the heavy mode is the Debye mass with mass $\sim g T$, and the light mode is spatial gauge and scalar fields. The purpose of DR is to build a model with only light mode, thus avoiding the IR problem that appeared in finite temperature field theory. In general, the DR needs two steps to obtain the thermal effective field theory containing only the light mode.

The original Lagrangian in this paper is Eq.\eqref{lagrangian}. The first step is to build a theory with heavy and light modes, that should integrate out fermion and non-zero boson modes.
With the accuracy $\mathcal{O}(g^4)$, the power counting we used in this paper are
$g^\prime\sim g, g_Y\sim g, \lambda\sim g^2, c_6\sim g^4/\Lambda^2$.
The $c_6$ term is non-renormalizable and contains the NP scale.
After integrating out superheavy mode, the effective theory  has the form
\begin{equation}\label{action3dheavy}
S^{\rmii{heavy}}_{\rmii{3d}} = \int {\rm d}^3 x \biggl[
      \frac{1}{4}G_{ij}^{a}G_{ij}^{a}
    + \frac{1}{4}F_{ij}^{ }F_{ij}^{ }
    +\frac{1}{2}(D_{i}^{ }A_{0}^{a})^2
    + \frac{1}{2}(\partial_{i}^{ }B_{0}^{ })^2
    + \frac{1}{2}(D_{i}^{ }C_{0}^{\alpha})^2
    + (D_{i}\Phi^{ })^\dagger (D_{i}\Phi^{ })
    + V^{\rmii{heavy}}_{\rmii{3d}}\biggr]
    \;,
\end{equation}
where
$G^{a}_{ij} =
\partial_{i} A^a_j -
\partial_{j} A^a_i + g_{3} \epsilon^{abc}A_{i}^b A_{j}^c$,
$F_{ij} = \partial_{i}B_{j} - \partial_{j}B_{i}$ and
$D_{i}\Phi = (\partial_{i} - ig_{3} \tau^a A_{i}^a/2 - ig^\prime_{3} B^{ }_{i}/2)\Phi^{ }$ with $\tau^a$ being the Pauli matrices. The heavy scalar potential has the form
\begin{equation}\label{eq:3d:soft:V}
\begin{aligned}
V^{\rmii{heavy}}_{\rmii{3d}} =&
    \mu_{h,3}^{2} \Phi^\dagger\Phi
    + \lambda^{ }_{3} (\Phi^\dagger\Phi)^2
    + c^{ }_{6,3} (\Phi^\dagger\Phi)^3
     \\ &
    + \frac{1}{2}\mD^2\,A^a_{0}A^a_{0}
    + \frac{1}{2}\mD'^2\,B_{0}^2
    + \frac{1}{2}\mD''^2\,C^{\alpha}_{0} C^{\alpha}_{0}
    \\ &
    + \frac{1}{4}\kappa_1 (A^a_{0}A^a_{0})^2
    + \frac{1}{4}\kappa_2 B_{0}^4
    + \frac{1}{4}\kappa_3 A^a_{0}A^a_{0}B_{0}^2
    \\ &
    + h^{ }_{1} \Phi^\dagger\Phi A^a_{0}A^a_{0}
    + h^{ }_{2} \Phi^\dagger\Phi B_0^2
    + h^{ }_{3} B_{0}\Phi^\dagger{A}_{0}^a\tau^a\Phi
    + h^{ }_{4} \Phi^\dagger\Phi C^\alpha_{0} C^\alpha_{0}
    \\&
    +\alpha_0 \epsilon_{ijk}\left(A_{i}^a G_{jk}^a-\frac{i}{3}g_3\epsilon^{abc} A_{i}^a A_{i}^b A_{k}^c\right)+\alpha^\prime\epsilon_{ijk} B_{i} F_{jk}
    \\&
    +\kappa_0 B_{0}+\rho \Phi^\dagger A_{0}^a\tau^a \Phi+\rho^\prime \Phi^\dagger \Phi B_{0}+\rho_G B_{0} A_{0}^a A_{0}^a
    \;,
\end{aligned}
\end{equation}
where $\mD,\mD',\mD''$ are the Debye masses of the time component of gauge fields.
The matching between 3d and 4d fields can be obtained by considering the momentum-dependent contribution of the superheavy modes to the 2-point Green function, and have the form
\begin{equation}\label{field}
\begin{aligned}
\phi_{3d}^2&=\frac{1}{T}Z_\phi \phi_{4d}^2\;,\\
A_{i,3d}^2&=\frac{1}{T}Z_{A_i} A_{i,4d}^2\;,\\
A_{0,3d}^2&=\frac{1}{T}Z_{A_0} A_{0,4d}^2\;,
\end{aligned}
\end{equation}
with
\begin{equation}\label{ZDR}
\begin{aligned}
Z_\phi^\prime&=1+\frac{1}{16\pi^2}\left(-\frac{3}{4}(3g^2+(g^\prime)^2 )L_b+3g_Y^2L_f\right)\;,\\ Z_\phi&=Z_\phi^\prime\left(1-\frac{3g_Y^2}{16\pi^2}\mathcal{A}\left(\frac{\mu_B}{3}\right)\right)\;,\\
Z_{A_i}^\prime&=1+\frac{g^2}{16\pi^2}\left(-\frac{25}{6}L_b-\frac{2}{3}+\frac{4}{3}N_F L_f\right)\;,\\ Z_{A_i}&=Z_{A_i}^\prime\left(1-\frac{g^2}{48\pi^2}\left[9\mathcal{A}\left(\frac{\mu_B}{3}\right)+\sum_{i=1}^{3}\mathcal{A}(\mu_{L_i})\right]\right)\;,\\
Z_{A_0}^\prime&=1+\frac{g^2}{16\pi^2}\left(-\frac{25}{6}L_b+3+\frac{4}{3}N_F (L_f-1)\right)\;,\\
Z_{A_0}&=Z_{A_0}^\prime\left(1-\frac{g^2}{48\pi^2}\left[9\mathcal{A}\left(\frac{\mu_B}{3}\right)+\sum_{i=1}^{3}\mathcal{A}(\mu_{L_i})\right]\right)\;,\\
\end{aligned}
\end{equation}
where the functions of $\mathcal{A}(x)$ describe the contribution of the chemical potential defined in the Eq.\eqref{functionA}\cite{Gynther:2003za}. Then, the relation between the 3d and 4d couplings can be obtained by vanishing external momenta contribution from superheavy mode to Green functions, the Debye masses can be obtained in the same way. Here, the Chern-Simons term $\alpha_0$ and $\alpha^\prime$ vanished by the assumption of Eq.~\ref{BLasum}.
See Appendix \ref{resdr} for these parameters formula.

The second step is to build a theory with only light mode. At the light scale, the fields $A_{0,3d}, B_{0,3d}$ and $C_{0,3d}$ are heavier and need to be integrated out as well. The action of light scale reads
\begin{equation}\label{eq:3d:ultrasoft:action}
S^{\rmii{light}}_{\rmii{3d}}= \int {\rm d}^3 x \biggl[
      \frac{1}{4}G_{ij}^{a}G_{ij}^{a}
    + \frac{1}{4}F_{ij}^{ }F_{ij}^{ }
    + (D_{i}\Phi^{ })^\dagger (D_{i}\Phi^{ })
    + V^{\rmii{light}}_{\rmii{3d}}\biggr]
    \;,
\end{equation}
with
\begin{equation}
\label{eq:3d:ultrasoft:V}
V^{\rmii{light}}_{\rmii{3d}} =\overline{\mu}_{h,3}^2\Phi^\dagger\Phi+\overline{\lambda}_3(\Phi^\dagger\Phi)^2+\overline{c}_{6,3}(\Phi^\dagger\Phi)^3
\end{equation}
and this theory is valid up to momenta $g^2T<p<gT$. There is no IR problem due to these fields are massive. We list the couplings of heavy scale eq.\eqref{action3dheavy} and light scale eq.\eqref{eq:3d:ultrasoft:action} in Appendix  \ref{resdr}.

Once we obtain the final light theory, the advantage of DR will be revealed. The light theory is simpler than the 4d theory with fermions and bosons whose masses are larger than $gT$ being integrated out. And all sum integrals have been evaluated. The temperature only enters the parameters of the light theory.

\subsection{The $3d$ thermal effective potential}

We construct the thermal 3d effective potential $V_{eff}$ using the light scale theory given by the Eq.\eqref{eq:3d:ultrasoft:action}, where the non-zero Matsubara modes and the temporal components of the gauge fields have been integrated out. Through the process, the renormalization scale and the chemical potential enter the mass parameters of $\bar{\mu}_{h,3}$ and the 3d couplings of $\bar{g}_3,\bar{g}^\prime_3$,$\bar{\lambda}_3$ and $\bar{c}_6$, See Appendix.\ref{appendixc} for details.
The effective potential up to the 2-loop order level has the same form as the SM, and the $(\Phi^\dagger\Phi)^3/\Lambda^2$ term leads to corrections to the parameters in the effective potential.
We use the $\hbar-$expansion to expand the thermal effective potential
\begin{equation}\label{potentialhb}
V_{eff}=V_{3(0)}+\hbar V_{3(1)}+\hbar^2 V_{3(2)}.
\end{equation}
After insert the background field to eq.\eqref{eq:3d:ultrasoft:V}, the tree-level potential is
\begin{align}\label{leading}
V_{3(0)} &=
    \frac{1}{2} \bar{\mu}_{h,3}^2 \phi_3^2
    + \frac{1}{4} \bar{\lambda}_3 \phi_3^4
    + \frac{1}{8} \bar{c}_{6,3} \phi_3^6
    \;.
\end{align}
The 1-loop level potential in the Landau gauge is
\begin{align}\label{oneloop}
V_{3(1)} =
    J_{\rmii{soft}}(m_{\phi,3})
    + 3 J_{\rmii{soft}}(m_{\chi,3})
    \nn
    + (d-1)\Big( 2 J_{\rmii{soft}}(m_{W,3}) + J_{\rmii{soft}}(m_{Z,3}) \Big)
\;,
\end{align}
with
\begin{align}
 J_{\rmii{soft}}(m) =
- \frac{1}{12\pi} (m^2)^\frac{3}{2}
\;.
\end{align}
Therein, the field dependent masses  are
\begin{align}
m^2_{\phi,3} &=
    \bar{\mu}_{h,3}^2
    + 3 \bar{\lambda}_3 \phi^2_3
    + \frac{15}{4} \bar{c}_{6,3} \phi^4_3
\;, \\
m^2_{\chi,3} &=
    \bar{\mu}_{h,3}^2
    + \bar{\lambda}_3  \phi^2_3
    + \frac{3}{4} \bar{c}_{6,3} \phi^4_3
\;,\\
m^2_{W,3} &=
    \frac{1}{4}\bar{g}_3^2 \phi^2_3
\;,\\
m^2_{Z,3} &=
    \frac{1}{4} \Big( \bar{g}_3^2
    + \bar{g}_3^{\prime 2} \Big) \phi^2_3
\;.
\end{align}
Since contribution from $\bar{c}_{6,3}$ appear through the scalar mass $m_{\chi,3}$ and $m_{\phi,3}$ and scalar self-interaction, the $\bar{c}_{6,3}$ term only affects (SS) and (SSS) part in the 2-loop potential of
\begin{align}
V_{3(2)} = -\Big(
    \text{(SSS)}
    + \text{(VSS)}
    + \text{(VVS)}
    + \text{(VVV)}
    + \text{(VGG)}
    + \text{(SS)}
    + \text{(VS)}
    + \text{(VV)}
    \Big)\;,
\end{align}
Different parts' contributions to the 2-loop effective potential are listed in Appendix.\ref{respotential}.

\subsection{Phase transition dynamics}\label{ptsub}
The Universe first lives in the ``symmetric'' phase, and as the Universe cools down, the ``symmetric'' and ``broken'' phases have the same free energy at the critical temperature, and it can obtained by solving
\begin{equation}
V_{eff}(\phi_c,T_c)=V_{eff}(0,T_c),\quad \left.\frac{\partial V_{eff}(\phi,T)}{\partial \phi}\right|_{\phi=\phi_c}=0\;.
\end{equation}
In particular, when the Universe cools further, the vacuum bubbles of the true phase nucleated in the symmetric phase.
 The nucleation temperature $T_n$ is obtained when the bubble nucleation rate $\Gamma=A \exp[-S_c]$ is equal to Hubble parameter $\Gamma\sim H$, i.e., $S_c\approx 140$~\cite{Linde:1981zj}.
 The Euclidean action is
\begin{equation}\label{action}
S_c=\int \bigg[\frac{1}{2}(\partial_i \phi)^2+V_{eff}(\phi,T) \bigg]d^3 x.
\end{equation}
The action can be obtained after the ``bounce solution" acquired
from the equations of motion:
\begin{equation}\label{bouncefun}
\frac{d^2\phi}{d\rho^2}+\frac{2}{\rho}\frac{d\phi}{d\rho}=\frac{dV_{eff}(\phi,T)}{d\phi}\;,
\end{equation}
with the boundary condition
\begin{equation}
\phi(\rho\rightarrow \infty)=0,\quad \left.\frac{d\phi}{d\rho}\right|_{\rho=0}=0\;.
\end{equation}
We use code ``findbounce'' to solve this equation and obtain the nucleation temperature $T_n$ and the corresponding background field value $\phi_n$~\cite{Guada:2020xnz}.
The inverse PT duration is defined as:
$\beta/H_n=T_n(dS_c/dT)|_{T_n}$.
The PT temperature and the duration determine the peak frequency of the produced GW from PT~\cite{Huber:2008hg,Caprini:2015zlo,Caprini:2009yp}, and the trace anomaly ($\alpha$) usually determines the amplitude of the generated GW. For the $4d$ theory, the $\alpha$ is defined as
$\alpha=\Delta\rho/\rho_{rad}$
with
\begin{equation}
\Delta\rho=\Delta V_{4d}(\phi_n,T_n)+\frac{1}{4}\left. T_n \frac{d \Delta V_{4d}(\phi_n,T)}{d T}\right|_{T=T_n}\;.\\
\end{equation}
After apply the relation between $4d$ and $3d$ potential $V_{4d}\approx T V_{eff}$, we have $\alpha=T (\Delta\rho/\rho_{rad})$ with
\begin{equation}
\Delta\rho=-\frac{3}{4}\Delta V_{eff}(\phi_n,T_n)+\frac{1}{4}\left. T_n \frac{d \Delta V_{eff}(\phi_n,T)}{d T}\right|_{T=T_n}\;,
\end{equation}
where
$
\Delta V_{eff}(\phi,T)=V_{eff}(\phi,T)-V_{eff}(0,T)
$ and $\rho_{rad}=\pi^2g_* T_n^4/30$, $g_*=106.75$ is the effective number of relativistic degrees of freedom~\cite{Croon:2020cgk}.

\subsection{Electroweak sphaleron}
In the electroweak theory, the sphaleron is a static and unstable solution of the field equations and corresponds to the top of the energy barrier between two topologically distinct vacua~\cite{Comelli:1999gt,Klinkhamer:1984di}.
The electroweak sphaleron process provides the baryon number violation condition (one of the three Sakharov conditions~\cite{Sakharov:1967dj}) to explain the baryon asymmetry of the Universe~\cite{Kuzmin:1985mm,Shaposhnikov:1987tw,Morrissey:2012db}. The electroweak sphaleron energy with $U(1)_Y$ contribution reads~\cite{Klinkhamer:1990fi,DeSimone:2011ek}:
\begin{equation}\label{spe3}
\begin{aligned}
E_{sph}^{EW}(\nu,T)=&\frac{4\pi v}{g}\frac{v(T)}{v}\int_{0}^{\infty}d\xi \left\{\sin^2{\nu}\left(\frac{8}{3}f^{\prime 2}+\frac{4}{3}f^{\prime 2}_3\right)+\sin^4{\nu}\frac{8}{\xi^2}\left[\frac{2}{3}f_3^2(1-f)^2\right.\right.\\
&\left.+\frac{1}{3}(f(1-f)+f-f_3)^2\right]+\frac{4}{3}\left(\frac{g}{g^\prime}\right)^2\left[\sin^2{\nu}f_0^{\prime 2}+\sin^4{\nu}\frac{2}{\xi^2}(1-f_0)^2\right]\\
&+\frac{1}{2}\xi^2h^{\prime 2}+\sin^2{\nu}h^2\left[\frac{1}{3}(f_0-f_3)^2+\frac{2}{3}(1-f)^2\right]+\left.\frac{\xi^2}{g^2 v(T)^4}V(h,T)\right\}\;,\\
\end{aligned}
\end{equation}
where $\nu$ is related to Chern-Simons number through $N_{CS}=\frac{2\nu-\sin{2\nu}}{2\pi}$.
This sphaleron energy is valid in $\nu\in[0,\pi]$ due to the $U(1)_Y$ gauge field will reduce the $SO(3)$ spherical symmetry to axial symmetry. Here, $h,f,f_0,f_3$ are functions that enter the Higgs and gauge field configurations, and $\xi$ is a dimensionless parameter with $\xi=g v r $, for more details see Appendix.~\ref{resspe}. In our work, we neglect the contribution of magnetic dipole moment. Here, we note that the gauge invariant Higgs condensate $\langle\Phi^\dag\Phi\rangle$ is related to the dimensionless quantity (the ratio between the Higgs vacuum expectation value (v) and the temperature) during the PT through $v/T=\sqrt{2\langle\Phi^\dag\Phi\rangle}/\sqrt{T}$, in this work we straightforwardly calculate the Higgs condensate as $\langle
\Phi^\dagger\Phi\rangle=d V_{eff}/d\overline{\mu}_{h,3}^2$ which applies when the Coleman-Weinberg type radiative correction is important, the spontaneous symmetry breaking is not necessarily to occur on the tree-level potential~\cite{Farakos:1994xh}.

 \section{First-order Phase transition parameters}\label{phasetransition}

  To study the renormalization scale dependence of the PT dynamics with the DR approach, we consider three energy scales for $\overline{\mu}$:
\begin{equation}
\begin{aligned}
\pi T\qquad &\text{the lowest fermionic mode}\nonumber\\
2\pi T\qquad &\text{the lowest nonzero bosonic mode}\nonumber\\
4\pi e^{-\gamma_E}T\qquad &\text{the lowest logarithmic contribution}\nonumber
\end{aligned}
\end{equation}
To demonstrate the effects from the chemical potential, we take $-\mu_B=\mu_{L_i}\equiv\mu_{ch}$ and consider two benchmark scenarios of $\mu_{ch}=0.1$T and   $\mu_{ch}=0.2$T.

\subsection{Phase transition temperatures}

\begin{figure}[!htp]
    \centering
      \includegraphics[width=0.4\textwidth]{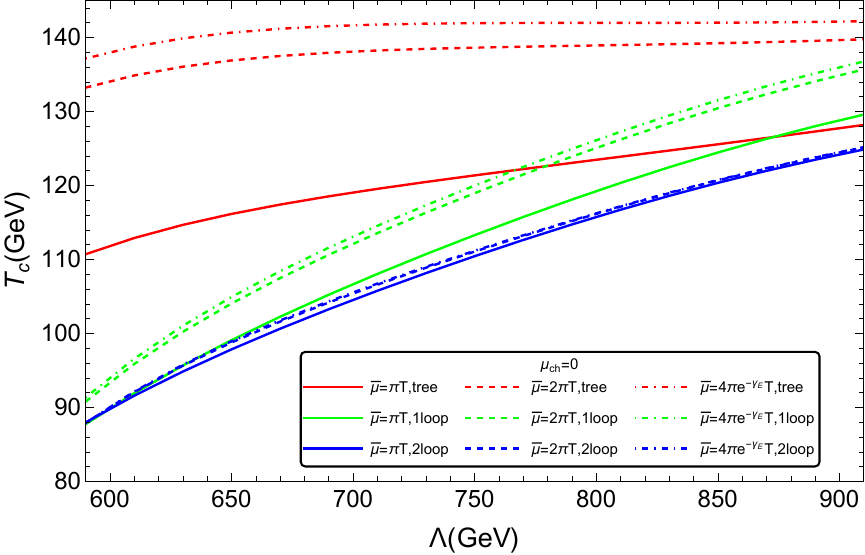}
    \includegraphics[width=0.4\textwidth]{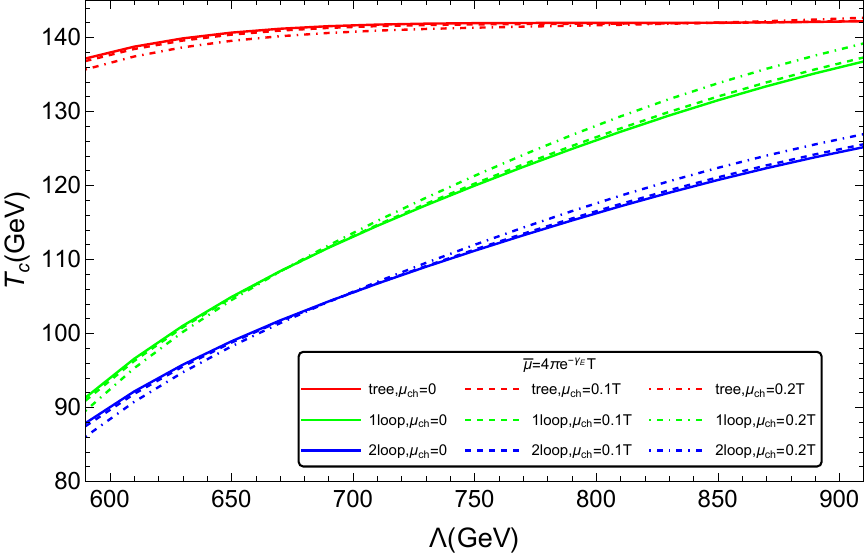}
    \includegraphics[width=0.4\textwidth]{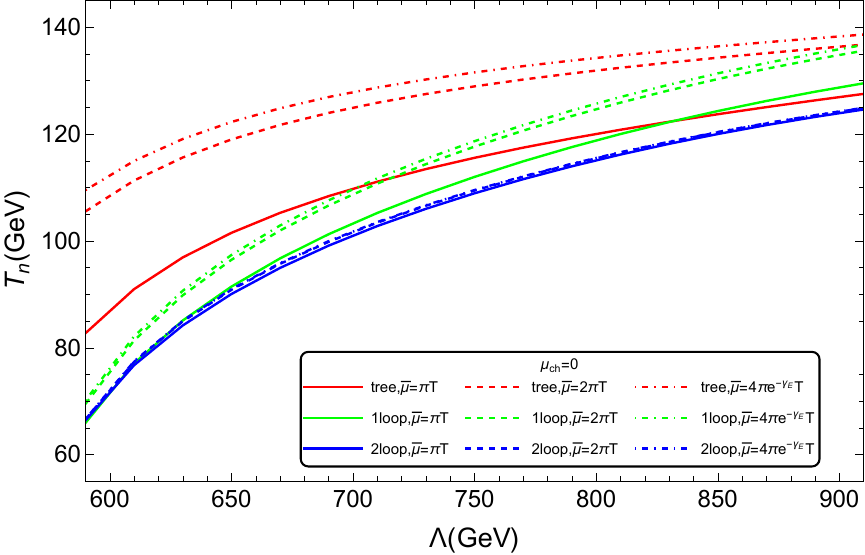}
    \includegraphics[width=0.4\textwidth]{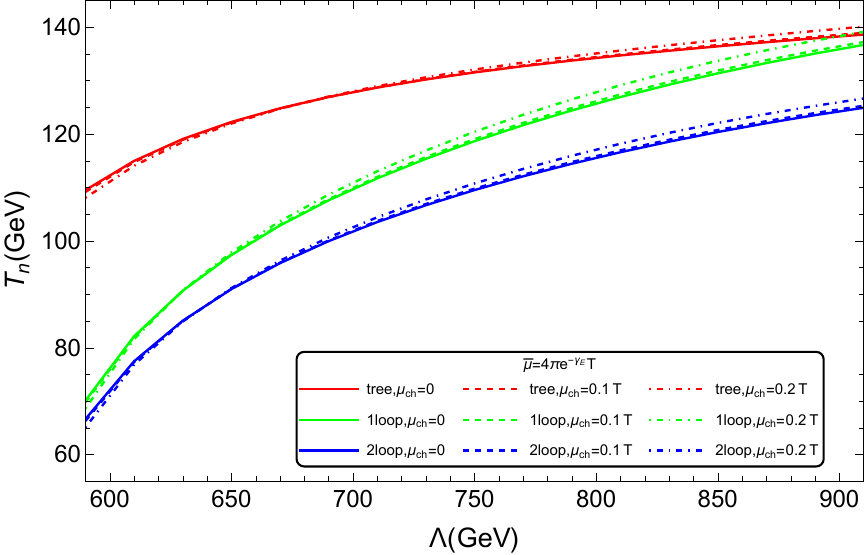}
   \caption{The renormalization scale dependence and the effects of the chemical potential in the critical temperature (upper panels) and nucleation temperature $T_n$ (lower panels). }
    \label{figtn}
\end{figure}

In Fig.\ref{figtn}, we present the renormalization scale dependence and the effect of the chemical potential on the critical temperature $T_c$ and the bubble nucleation temperature $T_n$ at different $\Lambda$ of the first-order PT. The renormalization scale dependence of $T_{c,n}$ at tree-level is very significant, and this dependence is reduced at higher loop orders, This scale dependence will lead to a large uncertainty, and make the result of tree-level unreliable. Generally, the magnitude of the $T_c$ and $T_n$ decrease as the calculation precision increases from one-loop to two-loop level. The magnitudes of the $T_n$ and $T_c$ increase with the increase of the NP scale, and decrease with the increase of precision of the calculation from tree-level to two-loop level. The renormalization scale dependence have $5\%$(one-loop) and $1\%$(two-loop) differences between the scenarios of renormalization scale at $\pi T$ and $4\pi e^{-\gamma_E}T$.
The effects of the chemical potential on $T_n$ and $T_c$ are much smaller relative to the renormalization scale dependence at tree- and one-loop levels. Since the renormalization scale dependence is significantly suppressed at the two-loop level, the effect of the chemical potential is relatively significant and needs to be considered there, see the right two plots of Fig.~\ref{figtn} where we set $\overline{\mu}=4\pi e^{-\gamma_E}T$ as an illustration. The nucleation temperature $T_n$ ranges from $65$ GeV to $125$ GeV at the two-loop level when the $\Lambda$ grows from 600 GeV to 900 GeV.
 The appearance of the chemical potential reduces (amplifies) the magnitude of the $T_c$ and $T_n$ at small (large) $\Lambda$.
 We observe around $1\% -2\%$ differences between the cases of $\mu_{ch}=0$ and $\mu_{ch}=0.2$T at each order at large $\Lambda$.

\subsection{Phase transition parameters for gravitational wave prediction}

\begin{figure}[!htp]
    \centering
    \includegraphics[width=0.4\textwidth]{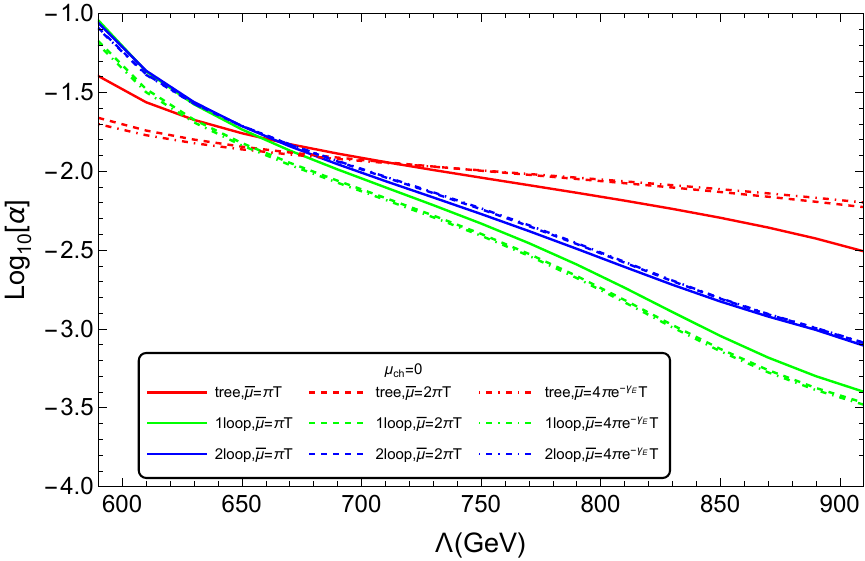}
    \includegraphics[width=0.4\textwidth]{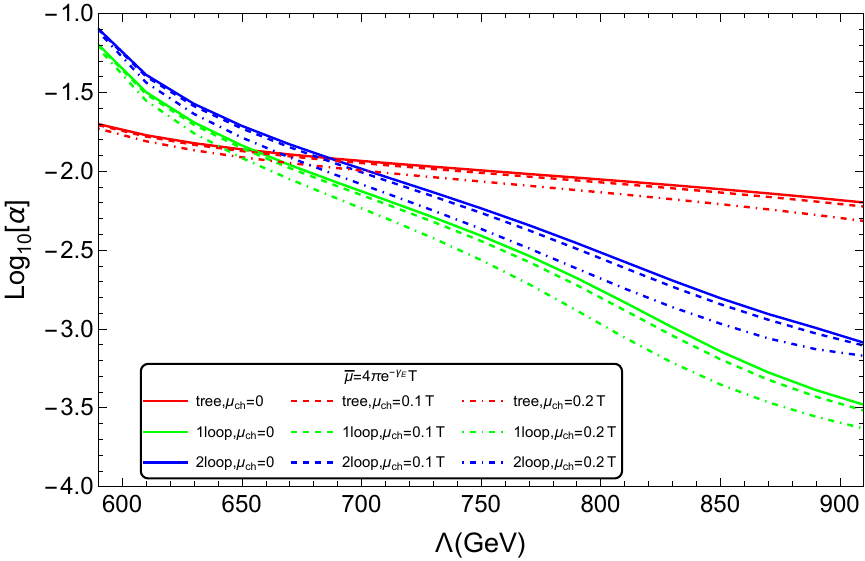}
    \caption{
    The renormalization scale dependence and chemical potential in the trace anomaly $\alpha$ for left and right plots.
 }
    \label{figalpha}
\end{figure}

In Fig.\ref{figalpha}, we illustrate the behaviors of the trace anomaly $\alpha$ at the tree-level and loop levels. When $\Lambda\sim 680$ GeV, the values of the $\alpha$ at tree level and loop levels are the same. When the $\Lambda$ is smaller, the magnitude of the $\alpha$ at tree level value will be larger than that of loop levels, and vice versa when $\Lambda\gtrsim 680$ GeV.
This phenomenon is because the contributions of loop level grow to be more important at
a larger NP scale when $\Lambda\gtrsim 680$ GeV, see Figs.(\ref{figpotential610},\ref{figpotential910}) for illustration.
Basically, this is because that the loop contributions of $V_{3(1,2)}$ would cancel the tree-level's contribution ($V_{3(0)}$) to the potential barrier $\Delta V_{eff}(\phi,T)$.
In the whole range of $\Lambda$, the effect of the renormalization scale dependence decreases with the increase of the precision of the calculation from the tree-level to the 2-loop level.
The chemical potential would affect the nucleation temperature and therefore can affect the minimal of the thermal effective potential, and in turn, affects the trace anomaly $\alpha$.
  This effect will be enlarged as the increase of the NP scale $\Lambda$.
  Our results show that the trace anomaly decreases with increasing chemical potential and that the effect of chemical potential is more important for weak PT (with small trace anomaly) at large $\Lambda$.
  The values of the trace anomaly ranges within the region of $\alpha\sim [\mathcal{O}(10^{-3})-\mathcal{O}(10^{-1})]$ for $\Lambda\sim[600,900]$ GeV, and the differences between the cases of $\mu_{ch}=0$ and $\mu_{ch}=0.2~T$ at 2-loop level increase from around $\sim 5\%$ to $\sim 35\%$
  as the $\Lambda$ grows.

\begin{figure}[!htp]
    \centering
    \includegraphics[width=0.4\textwidth]{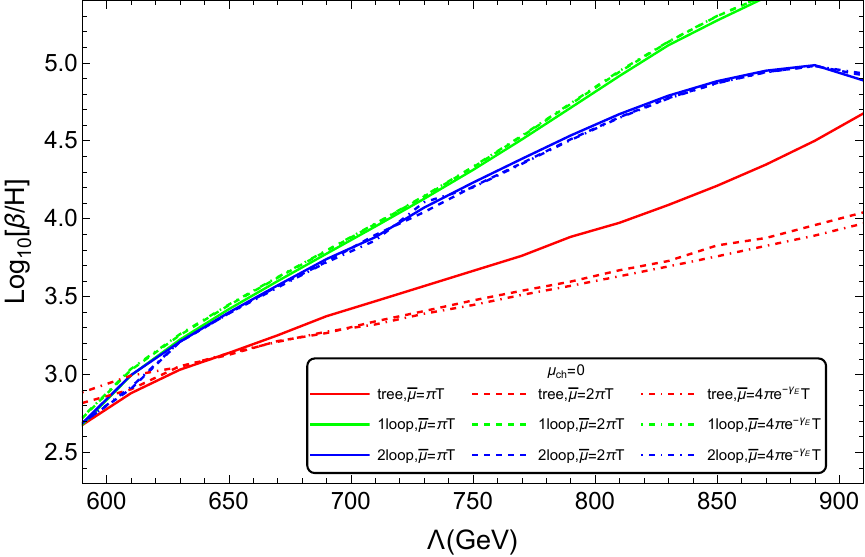}
\includegraphics[width=0.4\textwidth]{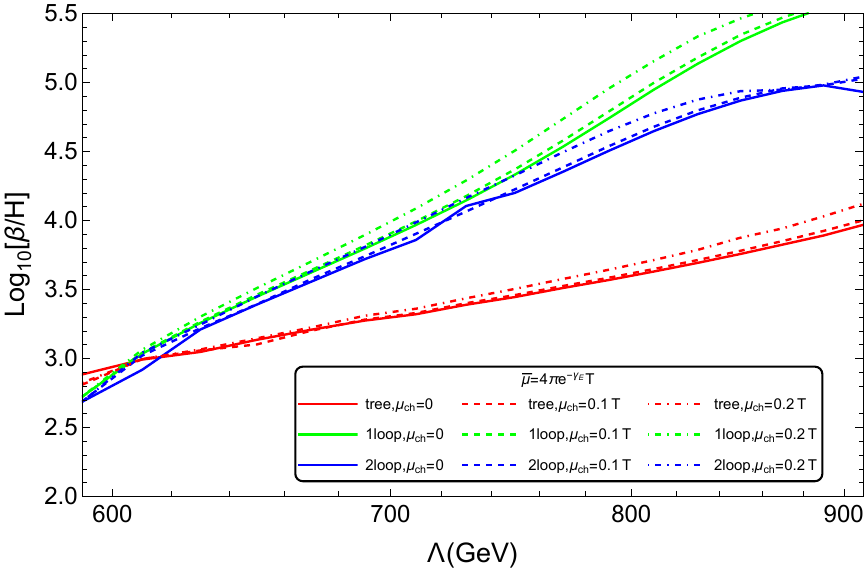}
    \caption{The magnitudes of $\beta/H_n$ as function of $\Lambda$. Left: The renormalization scale dependence on $\beta/H_n$. Right: The effect of chemical potential on $\beta/H_n$. }
    \label{figbetah}
\end{figure}

Figure.\ref{figbetah} shows how the $\beta/H_n$ depends on the renormalization scale dependence and the chemical potential. As in the PT temperatures of $T_c, T_n$ and the trace anomaly $\alpha$, the renormalization scale dependence of $\beta/H_n$ is magnificent at the tree-level, and the effect of the chemical potential grows as the NP scale increases. The difference between the value of $\beta/H_n$ at loop levels grows to be bigger and bigger than that of the tree-level as the $\Lambda$ increases. At small $\Lambda$ region, the values of one- and two-loop level of $\beta/H_n$ are relatively close to each other around $\mathcal{O}(10^2)-\mathcal{O}(10^4)$, while the difference between these two values is significant at large $\Lambda$ region. The $\beta/H_n$ is insensitive to renormalization scale dependence at loop levels, but there are $\sim37\%$ and $\sim 80\%$ differences between the renormalization scale at $\pi T$ and $4\pi e^{-\gamma_E}T$ at tree-level for large $\Lambda$ regions. The chemical potential has no significant effect on $\beta/H_n$ when $\Lambda$ is small, but it has an impact of about $20\%$ differences between the $\mu_{ch}=0$ and $\mu_{ch}=0.2T$ at large $\Lambda$ region at loop levels.

Recently, it was found that the $\alpha$ and $\beta/H$ are not independent parameters of first-order PT~\cite{Ellis:2020awk,Eichhorn:2020upj}. In the thin-wall limit, a
polynomial dependence of $\beta/H$ on $\alpha$ has been extracted as~\cite{Ellis:2020awk}:
\begin{equation}\label{thinwall}
\beta/H=T\frac{d}{dT}\frac{S_c}{T}\propto \frac{S_c}{T}\propto\alpha^{-2}.
\end{equation}

\begin{figure}[!htp]
    \centering
    \includegraphics[width=0.4\textwidth]{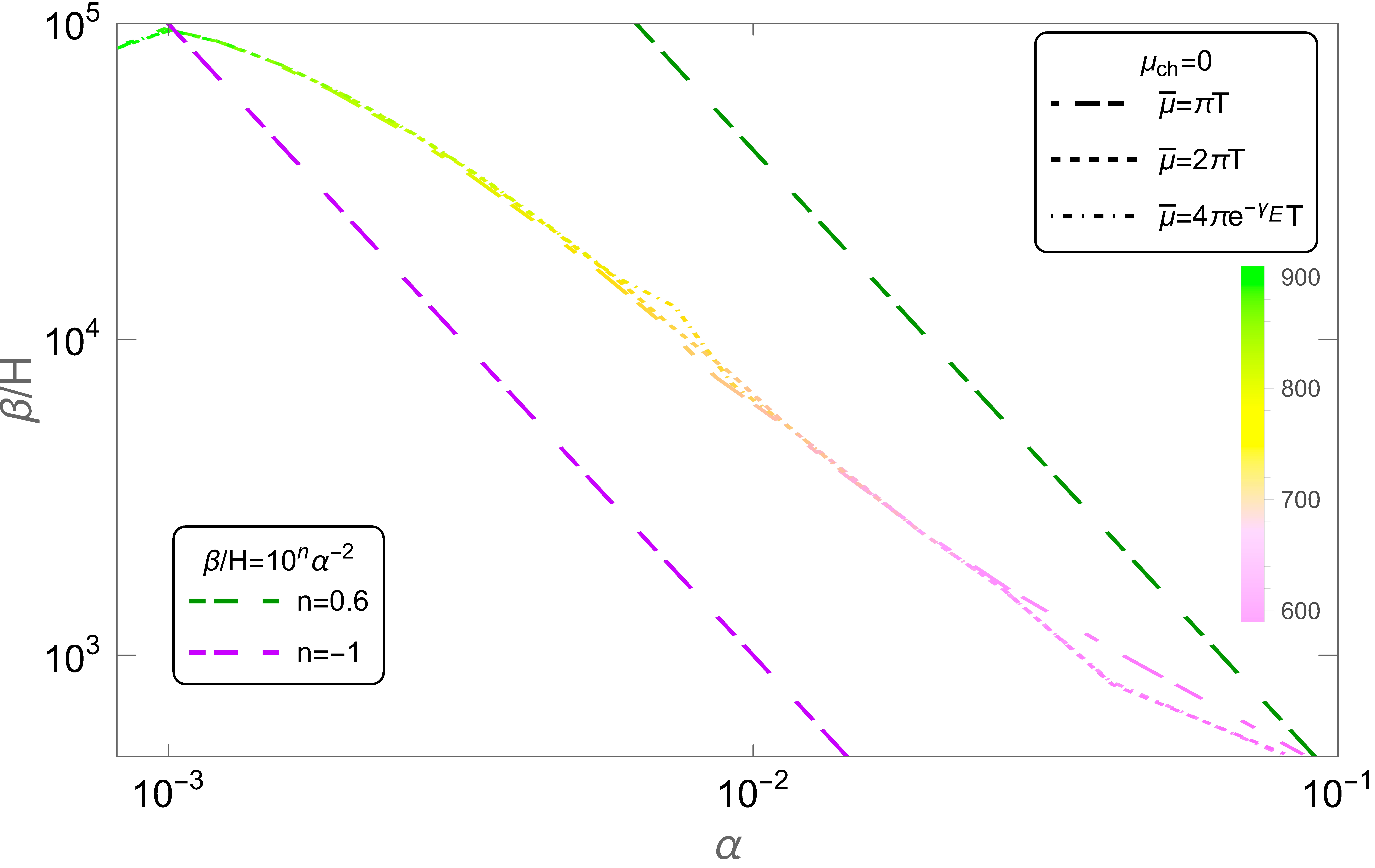}
    \includegraphics[width=0.4\textwidth]{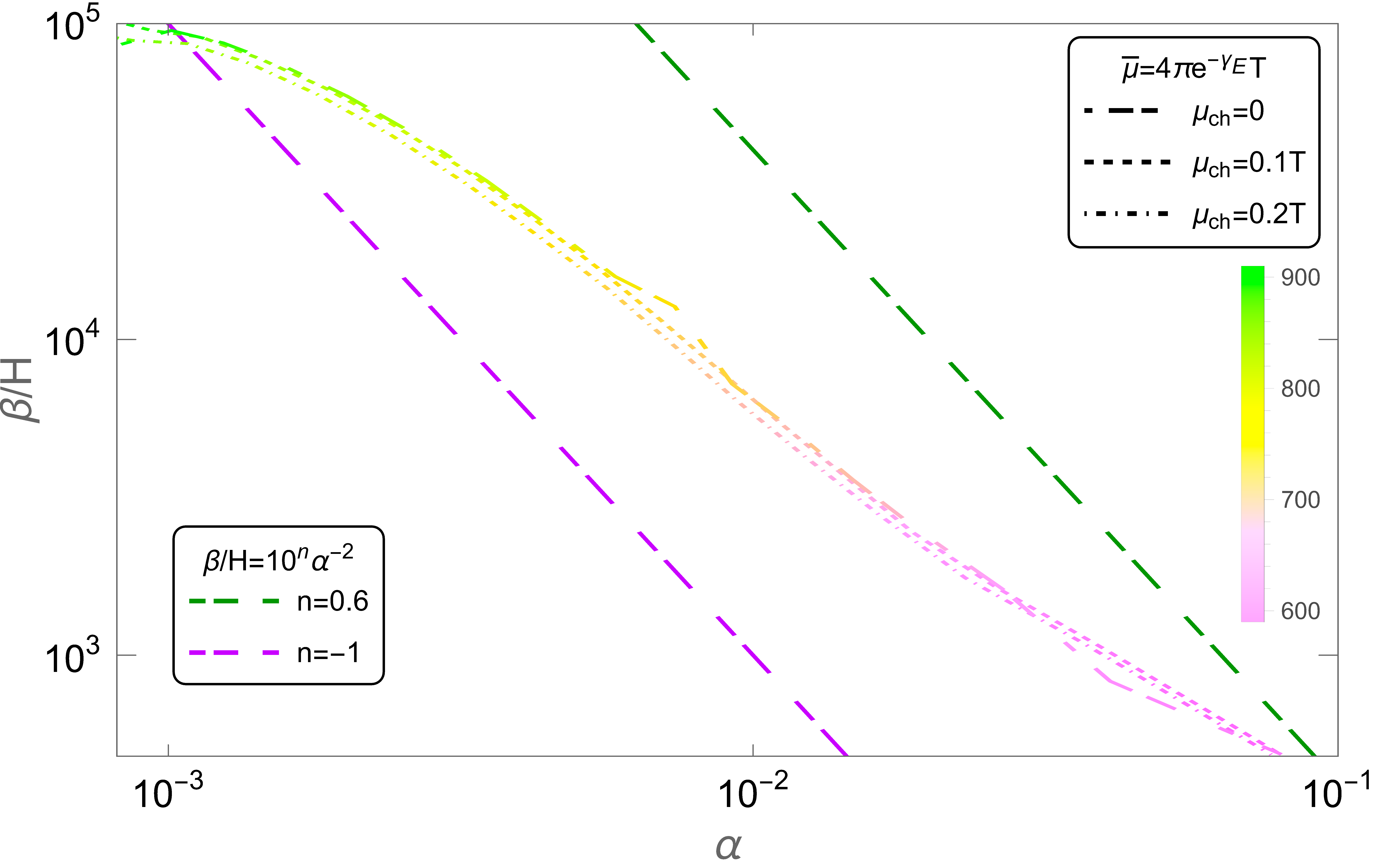}
    \caption{$\beta/H$ as function of $\alpha$. Left: the effect of renormalization scale dependence on the relation between $\beta/H$ and $\alpha$. Right: the effect of chemical potential on the association between $\beta/H$ and $\alpha$. The dashed lines correspond to the thin-wall approximation with a constant set to $10^{0.6}$(green) and $10^{-1}$(purple). The gradient color indicates the cut-off scale $\Lambda$.}
    \label{figthinwall}
\end{figure}

\begin{figure}[h]
    \centering
    \includegraphics[width=0.4\textwidth]{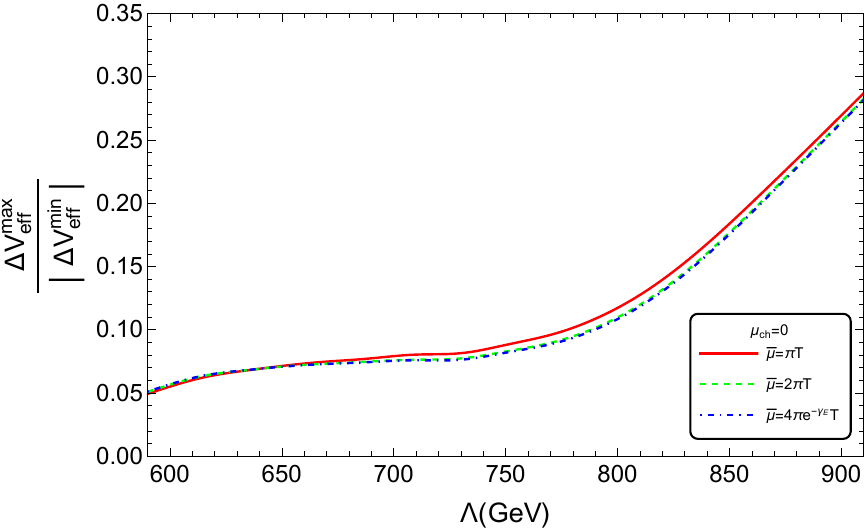}
    \includegraphics[width=0.4\textwidth]{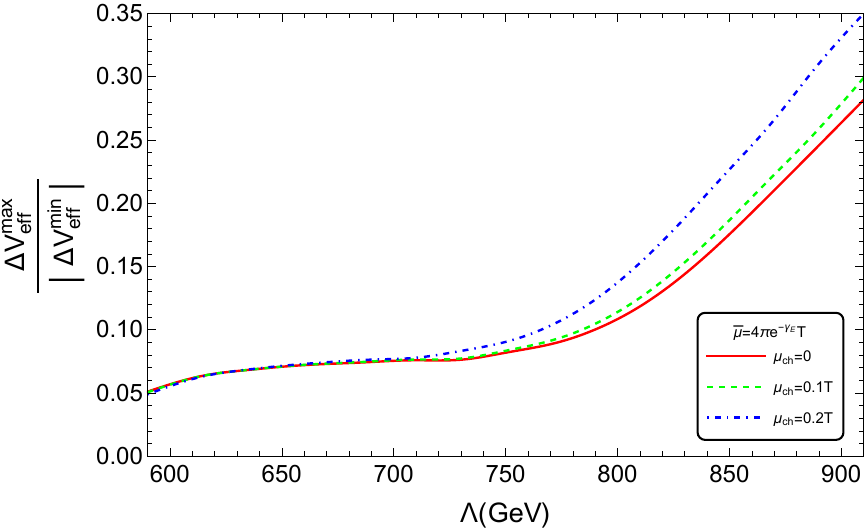}
    \caption{The ratio between the potential barrier height and the potential energy difference as a function of $\Lambda$ at two-loop order. These figures are plotted at the nucleation temperature $T_n$. Left: the renormalization scale dependence of ratio without considering chemical potential. Right: the effect of chemical potential on the ratio at $\overline{\mu}=4\pi e^{-\gamma_E }T$.}
    \label{figVmaxbyVmin}
\end{figure}

In Figure.\ref{figthinwall}, with the numerical results of mean bubble separation and the PT strength at two-loop level, we present the relation between the $\beta/H_n$ and $\alpha$, which shows that the prediction of Eq.~\ref{thinwall} is not really tenable here. We examine the
thin-wall limit in
Fig.\ref{figVmaxbyVmin}. It shows that the potential barrier height ($\Delta V_{eff}^{max}$) is smaller than the vacuum energy difference ($\Delta V_{eff}^{min}$), which makes the thin-wall approximation no longer applicable at the nucleation temperature $T_n$.

\subsection{Baryon number washout avoidance condition}\label{sphsection}

The sphaleron process will reach equilibrium at high temperature, and it should be strongly suppressed at low temperature to avoid washout of the generated baryon number which corresponds to the requirement of a strongly-first order electroweak PT~\cite{Patel:2011th}, i.e. $\Gamma_{sph}\varpropto \exp{(-E_{sph}^{EW}/T)}\ll H(T)$ with $E_{sph}^{EW}$ being the electroweak sphaleron energy,
this requires that the sphaleron rate in the broken phase is lower than the Hubble rate.
After reorganizing the relation among the sphaleron energy, PT temperature ($T_n$), and the scalar VEV, the baryon number washout avoidance condition can be obtained to restrict the baryon asymmetry explanation parameter spaces~\cite{Zhou:2019uzq,Gan:2017mcv}:
\begin{equation}\label{PT2}
PT_{sph}\equiv\frac{E_{sph}}{T}-7\ln\frac{v}{T}+\ln\frac{T_n}{100\text{GeV}}>(35.9-42.8)\;,
\end{equation}

\begin{figure}[!htp]
   \centering     \includegraphics[width=0.4\textwidth]{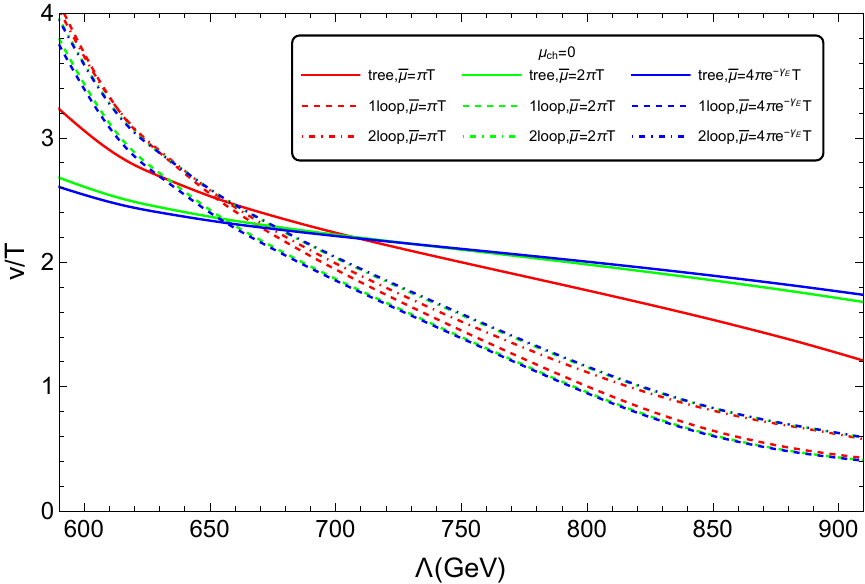}    \includegraphics[width=0.4\textwidth]{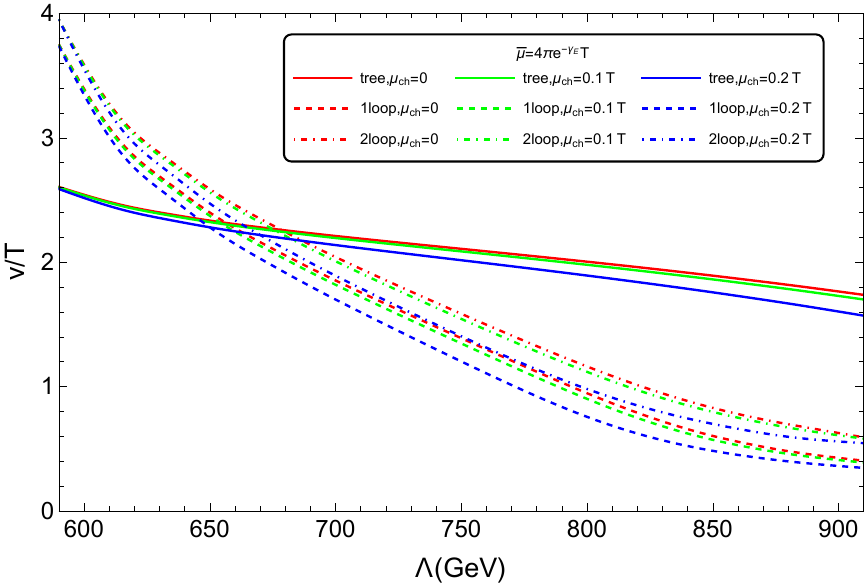}
   \caption{The quantity $v/T$ as a function of $\Lambda$. Left: the renormalization scale dependence of $v/T$ at $T_n$. Right: the effect of chemical potential on $v/T$ with $\overline{\mu}=4\pi e^{-\gamma_E}T$ at $T_n$.}
    \label{figvbyT}
\end{figure}

In the two plots of Fig.~\ref{figvbyT}, we present the values of $v/T$ at loop levels that are dramatically different from that of the tree-level. The so-called {\it strong first-order PT condition} in literatures $v/T\geq 1$ is stasified in all the regions of $\Lambda$ under study, while $v/T\leq 1$ when $\Lambda\gtrsim 800$ GeV at loop levels.
The bottom-left plot illustrates the renormalization scale dependence at loop levels is weaker than at tree level. The right one demonstrates the effect of the chemical potential is stronger at large $\Lambda$ region.
These two plots show that, at loop levels, the effect of the chemical potential is slightly more pronounced compared to the renormalization scale dependence in the large $\Lambda$ region.

\begin{figure}[!htp]
    \centering \includegraphics[width=0.4\textwidth]{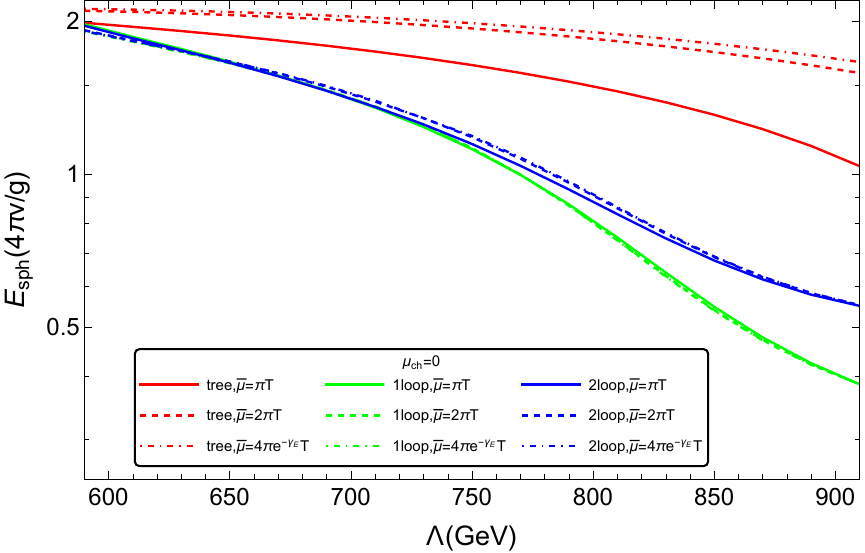}
    \includegraphics[width=0.4\textwidth]{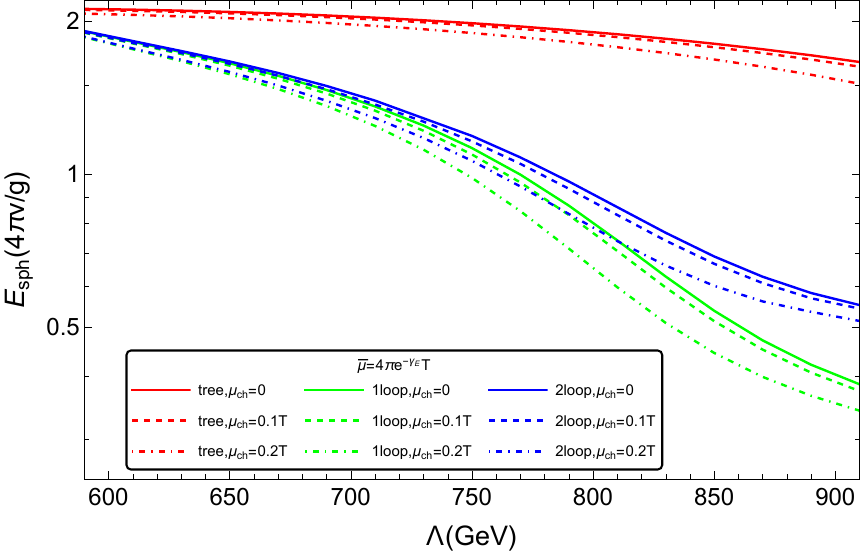}
    \caption{The renormalization scale dependence and the effects of the chemical potential in the sphaleron energy at different NP scales $\Lambda$.}
    \label{figsp}
\end{figure}

In the figure.\ref{figsp}, the renormalization scale dependence at 2-loop level is very weak and can be ignored.
The sphaleron energy at 2-loop grows to be larger than that at 1-loop level after $\Lambda\gtrsim 700$ GeV.
The chemical potential reduces the sphaleron energy to be lower, and this effect at large $\Lambda$ is stronger than the case with small $\Lambda$. This effect is caused by the effect of chemical potential on the thermal effective potential, see Figs.(\ref{figpotential610}, \ref{figpotential910}) in Appendix~\ref{respotential} for illustration where the chemical potential significantly changes the location of the minimum value of the potential energy at large $\Lambda$.
At the same time, we want to point out that the effect of the chemical potential on the sphaleron solution is negligible, see Fig.~\ref{resspe} in Appendix~\ref{respotential} for an illustration.

\begin{figure}[!htp]
  \centering
\includegraphics[width=0.4\textwidth]{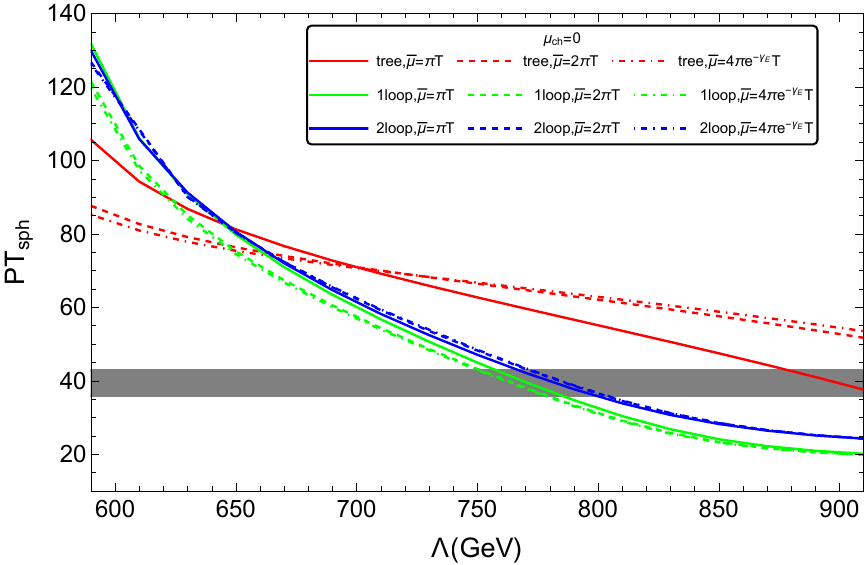}
\includegraphics[width=0.4\textwidth]{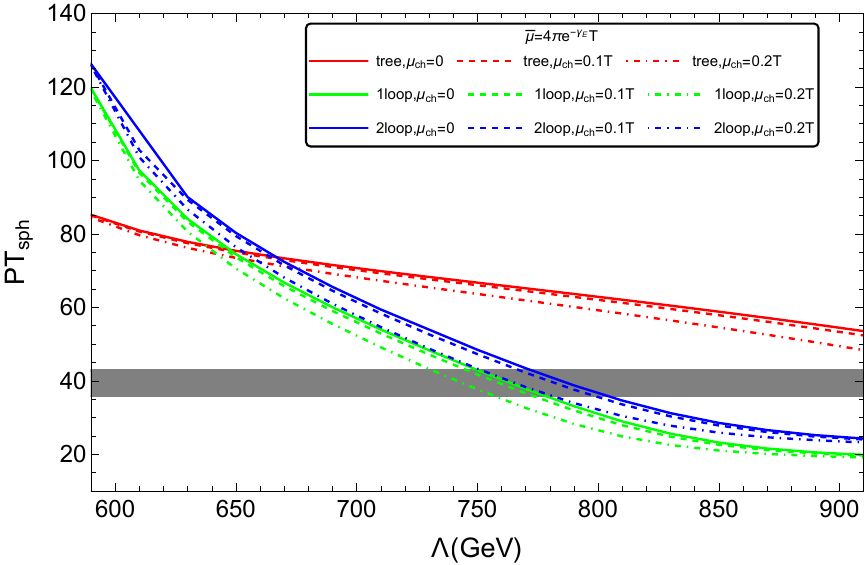}
  \caption{The quantity $PT_{sph}$ as a function of $\Lambda$. Left: the renormalization scale dependence of $PT_{sph}$ with $\mu_{ch}=0$. Right: the effect of chemical potential on $PT_{sph}$ with $\overline{\mu}=4\pi e^{-\gamma_E}T$.}\label{figpt}
\end{figure}

Fig.\ref{figpt} shows the renormalization scale dependence and the effect of chemical potential on the behavior of the quantity $PT_{sph}$.
The gray region represents the uncertainty in Eq.\eqref{PT2}. Above the gray region, the sphaleron process is sufficiently suppressed to satisfy the requirement that the sphaleron rate be lower than the Hubble rate.
The two plots demonstrate that the higher the NP scale $\Lambda$, the more pronounced the effects of the renormalization scale and chemical potential dependences.
As shown in the left panel, the renormalization scale dependence of $PT_{sph}$ at loop levels is weaker than that of the tree level. Comparatively, at the loop levels, the effect of the chemical potential on the $PT_{sph}$ is stronger than the theoretical uncertainty caused by the renormalization scale dependence, but the opposite is true at the tree level.
The increase of the chemical potential yields the decrease of the value of $PT_{sph}$ since the magnitude of the $E_{sph}/T$ decreases as $\mu_{ch}$ increases and $\ln (v/T)$ is not so sensitive to $\mu_{ch}$. This leads to more tight constraints on the NP scale.

In this study, at the 2-loop level, the baryon number washout avoidance condition requires the NP scale to be $\Lambda\lesssim (750-780)$ GeV for $\mu_{ch}=0.2$ T and $\Lambda\lesssim (770-800)$ GeV for $\mu_{ch}=0$. Meanwhile, the strongly first-order PT condition yields $\Lambda\lesssim 800 (830)$ GeV for the scenario without (with) considering the effect of the chemical potential.
For the case without considering the chemical potential, the washout avoidance condition yields slightly looser constraints on the NP scale $\Lambda$ than that of  Ref.\cite{Zhou:2019uzq}. This is mainly caused by the corrections of loop levels.

\section{Conclusion}\label{endsection}

In this paper, we have studied the renormalization scale dependence and the chemical potential effect of phase transition, and the strong first-order PT through the dimensional reduction approach. The results of this paper are that the dimensional reduction at 2-loop level reduces the theoretical uncertainty caused by renormalization scale dependence very well. The chemical potential has a greater effect on the weak PT region than it does on the strong PT region.

We find that when the NP scale $\Lambda\gtrsim 680$ GeV, the loop level contributions become more and more important for the thermal effective potential.
The renormalization scale dependence affects the thermal effective potential more significantly than that of PT temperature $T$, and the trace anomaly (or the PT strength) $\alpha$ is very significantly affected by the renormalization scale dependence.
The effect of the chemical potential on the PT temperature is different at different ranges of $\Lambda$. At small $\Lambda$, the PT temperature increases with increasing chemical potential in most first-order PT regions of $\Lambda$.
The chemical potential will reduce the PT strength and sphaleron energy to be lower. The chemical potential affects the baryon number washout avoidance condition mainly through the sphaleron energy, and its effect becomes evident in the weak PT region at large NP scales.
This phenomenon becomes more pronounced as $\Lambda$ increases due to the effect of the dimension-six operator becomes weaker and the effect of chemical potential becomes more prominent.

Our calculations suggest that, at two-loop level, the nucleation temperature $T_n$ would be lower than that of the tree-level around $\mathcal{O}(1-10)$ GeV, and the inverse PT duration grows several times than that of the tree-level. Therefore, the peak frequency of the predicted gravitational wave would be shifted to more higher frequency from tree-level to loop levels since the peak frequency is proportional to the value of $\beta/H$.
Our study also shows that the PT strength would be greater (smaller) than that of the tree-level when $\Lambda$ is smaller(greater) than $\sim 680$ GeV, therefore the predicted GW signal would be suppressed (amplified) since the amplitude of the GW from PT is mostly characterized by the PT strength. In comparison with the strong first-order PT condition, the baryon number washout avoidance condition yields more stringent constraints on the NP scale, with $\Lambda\lesssim (770-800)$ GeV at two-loop level, and the constraints will become more rigorous when the chemical potential is taken into account.

\section{Acknowledgements}

This work is supported by the National Key Research and Development Program of China under Grant No. 2021YFC2203004.
L.B. is supported by the National Natural Science Foundation of China (NSFC) under Grants Nos. 12075041, 12147102, 12322505. L.B. also acknowledges Chongqing Talents: Exceptional Young Talents Project No. cstc2024ycjh-bgzxm0020..

\appendix

\section{The realtions between $\overline{MS}$ parameters and physical observables}\label{Parameter}
We use the dimensional regularisation in the $\overline{MS}-$scheme in this work. In the Landau gauge, the $\beta-$functions at one-loop order read
\begin{align}
\overline{\mu}\frac{d}{d\overline{\mu}}\mu_h^2&=\frac{1}{(4\pi)^2}\mu_h^2\left(-\frac{3}{4}(3g^2+g^{\prime 2})+12\lambda+6g_Y^2\right),\\
\overline{\mu}\frac{d}{d\overline{\mu}}\lambda&=\frac{1}{(4\pi)^2}\left(\frac{3}{8}(3g^4+g^{\prime 4}+2g^2 g^{\prime 2})+24\lambda^2-6g_Y^4+24c_6\mu_h^2-3\lambda(3g^2+g^{\prime 2}-4g_Y^2)\right),\\
\overline{\mu}\frac{d}{d\overline{\mu}}c_6&=\frac{1}{(4\pi)^2}c_6(108\lambda-\frac{9}{2}(3g^2+g^{\prime 2})+18g_Y^2).
\end{align}
With these beta functions, one can obtain these couplings at different energy scales that will appear in the 3d parameters after integrating out the heavy and super-heavy modes.
We relate the $\overline{MS}-$parameters to physical observables by input Fermi constant $G_F=1.1663787\times 10^{-5}\text{GeV}^{-2}$ and pole mass\cite{ParticleDataGroup:2022pth}
\begin{equation}\label{massparameter1}
(M_t,M_W,M_Z,M_h)=(172.76,80.379,91.1876,125.1)~\text{GeV}.
\end{equation}
After use the shorthand notation $\g_0^2=4\sqrt{2}G_\mu M_W^2$, the tree level relation for gauge and Yukawa couplings($g_Y$) are read
\begin{equation}\label{coulpingparameter1}
g^2=g_0^2,\quad g^{\prime 2}=g_0^2\left(\frac{M_Z^2}{M_W^2}-1\right),\quad g_Y=\sqrt{\frac{1}{2} g_0^2 \frac{M_t^2}{m_W^2}}.
\end{equation}
and $v_0^2=4M_W^2/g_0^2=(246.22\text{GeV})^2$ for the tree level VEV. We set the strong coupling constant $g_s^2=4\pi\alpha_s(m_z)$ with $\alpha_s(m_z)=0.1180$ and neglect its runing\cite{ParticleDataGroup:2022pth}.

The tree-level Higgs mass parameter and self-interaction can be obtained by solving the
\begin{equation}
\left.\frac{\partial^2 V_0}{\partial \phi^2}\right|_{\phi=v_0}=M_h^2,\quad \left.\frac{\partial V_0}{\partial\phi}\right|_{\phi=v_0}=0
\end{equation}
where $V_0$ is defined in eq.\eqref{leading}. Then we can obtain
\begin{equation}
\mu_h^2=-\frac{1}{2}M_h^2+\frac{3}{4}c_6v_0^4,\quad  \lambda=\frac{1}{2}\frac{M_h^2}{v_0^2}-\frac{3}{2}c_6 v_0^2.
\end{equation}

\begin{figure}[!htp]
  \centering  \includegraphics[width=0.4\textwidth]{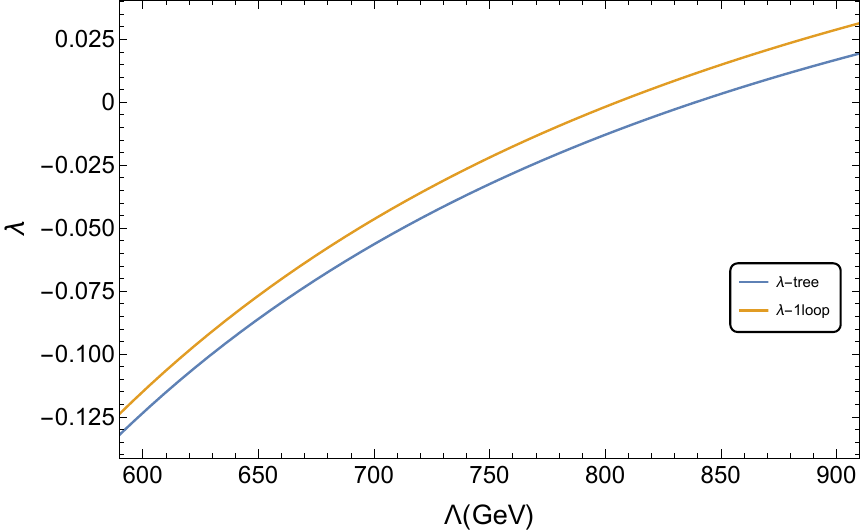}
  \includegraphics[width=0.4\textwidth]{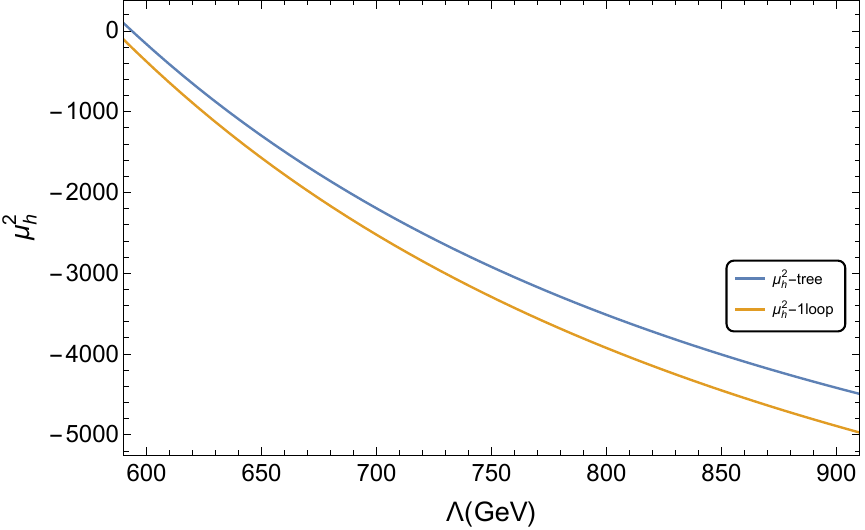}
  \caption{The scalar quartic coupling and the Higgs mass parameter as functions of the NP scale $\Lambda$.}\label{figpara}
\end{figure}

The tree-level relation is at $\mathcal{O}(g^2)$ accuracy, for the complete $\mathcal{O}(g^4)$ accuracy of the 2-loop order, the tree level relations need to be improved by their one-loop corrections. These corrections have been done in Ref.\cite{Croon:2020cgk}, and we show the result in Fig.\ref{figpara} for the tree-level relation and its one-loop correction. We would like to mention that, in our calculations, the $\lambda$ is different from that of Ref.\cite{Croon:2020cgk} which directly affects the thermal effective potential, and the PT temperature will be increased when the scalar quartic coupling $\lambda$ is decreased.
For example, at tree level, the $\lambda=-0.12$ with $\Lambda=600~\text{GeV}$ while $\lambda=-0.16$ with $\Lambda=600~\text{GeV}$ in Fig.11 of Ref.\cite{Croon:2020cgk}.
Meanwhile, the behavior of
$\mu_h^2$ is the same as that of Ref.\cite{Croon:2020cgk}. Our results of $T_c,T_n, \alpha,\beta/H$ therefore differ from that of Ref.\cite{Croon:2020cgk}.
However, our results also show that the renormalization scale dependence of these physical quantities will be lower at the loop level than the tree level, which is the same as the results in Ref.\cite{Croon:2020cgk}.

\section{Parameters of dimensional reduction}\label{resdr}\label{appendixc}

We use the code ``DRalgo'' to obtain the parameters of dimensional reduction~\cite{Ekstedt:2022bff}. The tree and one-loop result of ``DRalgo'' can be directly matched with literature, but for the two-loop level result, the ``DRalgo'' uses the Glaisher-Kinkelin constant. The relation between the Glaisher-Kinkelin constant $A$ and Riemann zeta function $\zeta(x)$ is
\begin{equation}
 \ln(2\pi)-(\ln(\zeta(2)))^\prime =1-\gammaE+(\ln(\zeta(-1))^\prime,\quad
 1+(\ln(\zeta(-1))^\prime =12\ln[A]\;,
\end{equation}
where $\ln[A]=(1/12)\left(\gammaE-L_b-2\left(c+\log\left[3T/\bmu\right]\right)\right)\;$,
with
\begin{align}
c \equiv
    \frac{1}{2}\bigg(\ln\Big(\frac{8\pi}{9}\Big) + \frac{\zeta'(2)}{\zeta(2)} - 2 \gammaE \bigg)
    \;,
\end{align}
  and
\begin{align}
\label{eq:Lb}
    L_b &\equiv2\ln\Big(\frac{\bmu}{T}\Big)-2[\ln(4\pi)-\gammaE]
    \;,\\
\label{eq:Lf}
    L_f &\equiv L_b + 4\ln2
    \;,
\end{align}
$\gammaE$ is the Euler-Mascheroni constant.
It should be noticed that ``DRalgo'' is worked in Landau gauge.
The chemical potential part of the parameters has been calculated in Ref.\cite{Gynther:2003za}, we add the chemical potential part contributions to the corresponding parameters. After defining the function\cite{Gynther:2003za}
{\setlength{\abovedisplayskip}{5pt}
\setlength{\belowdisplayskip}{5pt}
\begin{equation}\label{functionA}
\mathcal{A}(\mu_{ch})=\psi(\frac{1}{2}+\frac{i\mu_{ch}}{2\pi T})+\psi(\frac{1}{2}-\frac{i\mu_{ch}}{2\pi T})+2\gammaE+2\ln{4},\quad \psi(z)=\partial_z \ln{\Gamma(z)}
\end{equation}}
the final results in Landau gauge ($ \xi=0$) for the 3d parameters at the heavy scale read

\begin{align}
\lambda_{3} =&
    T\Big(\lambda(\bmu) + \frac{1}{(4\pi)^2}\bigg[\frac{1}{8}\Big(3\g^{4} + {\gp}^{4} +2 \g^{2} {\gp}^2 \Big)+ 3 L_f \Big(\gY^{4} - 2\lambda \gY^{2} \Big)\nn\\ &
    - L_b \bigg(\frac{3}{16}\Big(3\g^{4} + {\gp}^{4} + 2 \g^{2} {\gp}^{2} \Big) -\frac{3}{2}\Big(3\g^{2}+{\gp}^{2} -8 \lambda \Big) \lambda \bigg) \bigg]\Big)
    \nn \\ &
    + T^3 c_6
    - \frac{\mh^2}{(4\pi)^2} 12 T c_{6}L_{b}-\frac{3g_Y^2 T}{16\pi^2}(g_Y^2-2\lambda)\mathcal{A}\left(\frac{\mu_B}{3}\right)
    \;, \\
\label{eq:c63}
c_{6,3} =&
    T^2 c_6(\bmu) \bigg (1 + \frac{1}{(4\pi)^2} \bigg[ \Big( -54 \lambda + \frac{9}{4} (3 \g^{2} + {\gp}^{2}) \Big) L_b- 9 \gY^{2} L_f \bigg] \bigg)
    \nn \\ &
    - \frac{\zeta(3)}{768 \pi^4} \bigg( -\frac{3}{8} \Big( 3 \g^{6} + 3 \g^{4} {\gp}^{2} + 3 \g^{2} {\gp}^{4} + {\gp}^{6} \Big)  - 240 \lambda^3
 +84 \gY^{6}\bigg)
    \;, \\
\g_{3}^2 =&
    \g^{2}(\bmu)T\bigg[1 +\frac{\g^{2}}{(4\pi)^2}\bigg(\frac{43}{6}L_b+\frac{2}{3}-\frac{4\Nf}{3}L_f\bigg)\bigg]+\frac{g^4 T}{48\pi^2}\left[9\mathcal{A}\left(\frac{\mu_B}{3}\right)+\sum_{i=1}^3\mathcal{A}(\mu_{L_i})\right]
    \;,\\
{\gp_{3}}^{2} =&
    {\gp}^{2}(\bmu)T\bigg[1 +\frac{{\gp}^{2}}{(4\pi)^2}\bigg(-\frac{1}{6}L_b-\frac{20\Nf}{9}L_f\bigg)\bigg]
    \;,
            \end{align}

\begin{align}
h_{1} =&
    \frac{\g^{2}(\bmu)T}{4}\bigg(1+\frac{1}{(4\pi)^2}\bigg\{\bigg[\frac{43}{6}L_b+\frac{17}{2}-\frac{4\Nf}{3}(L_f-1)\bigg]\g^{2}+\frac{{\gp}^{2}}{2}-6\gY^{2}+12\lambda\bigg\} \bigg)\nn\\
    &+\frac{g^4 T}{192\pi^2}\left[9\mathcal{A}\left(\frac{\mu_B}{3}\right)+\sum_{i=1}^3\mathcal{A}(\mu_{L_i})\right]
    \;,\\
h_{2} =&
    \frac{{\gp}^{2}(\bmu)T}{4}\bigg(1 +\frac{1}{(4\pi)^2}\bigg\{\frac{3\g^{2}}{2}
    - \bigg[\frac{(L_b-1)}{6}  + \frac{20\Nf(L_f-1)}{9}\bigg]{\gp}^{2}- \frac{34}{3} \gY^{2}+12\lambda\bigg\} \bigg)
    \;,\\
h_{3} =&
    \frac{\g(\bmu){\gp}(\bmu)T}{2}\bigg\{1+\frac{1}{(4\pi)^2}\bigg[-\g^{2}+ \frac{1}{3}{\gp}^{2}+L_b\bigg(\frac{43}{12}\g^{2} -\frac{1}{12}{\gp}^{2}\bigg)\nn \\ &
    - \Nf(L_f-1)\bigg(\frac{2}{3}\g^{2}+\frac{10}{9}{\gp}^{2}\bigg)+4\lambda+ 2\gY^{2}\bigg] \bigg\}
    \;, \\
h_{4} =& -T\frac{1}{(4\pi)^2} 2 \gs^{2} \gY^{2}
    \;, \\
\kappa_{1} =& T\frac{\g^{4}}{(4\pi)^2} \bigg(\frac{17-4\Nf}{3}\bigg)
    \;,\\
\kappa_{2} =& T\frac{{\gp}^{4}}{(4\pi)^2} \bigg(\frac{1}{3}-\frac{380}{81} \Nf\bigg)
    \;,\\
\kappa_{3} =& T\frac{\g^{2} {\gp}^{2}}{(4\pi)^2}\bigg(2-\frac{8}{3}\Nf\bigg)
    \;, \\
\mD^{2} =&
    \g^{2}(\bmu) T^2\bigg(\frac{5}{6}+\frac{\Nf}{3}\bigg)+\frac{g^2}{4\pi^2}\left(\mu_B^2+\sum_{i=1}^3\mu_{L_i}^2\right)
    \;,\\
\mD'^{2} =&
    {\gp}^{2}(\bmu)  T^2\bigg(\frac{1}{6}+\frac{5\Nf}{9}\bigg)+\frac{g^{\prime 2}}{4\pi^2}\left(\frac{11}{9}\mu_B^2+3\sum_{i=1}^3\mu_{L_i}^2\right)
    \;,\\
\mD''^{2} =&
    \gs^{2} T^2 \bigg(1+\frac{\Nf}{3}\bigg)
    \;,\\
\mu_{h,3}^2 =&
    \mu^2_{h}(\bmu)
    + \frac{T^2}{16}\Big(3\g^{2}(\bmu) + {\gp}^{2}(\bmu) + 4 \gY^{2}(\bmu) + 8 \lambda(\bmu) \Big)
    + \frac{1}{4} T^4 c_6
    \nn \\ &
    + \frac{1}{(4\pi)^2} \bigg\{ \mh^2 \bigg[ \Big(\frac{3}{4}(3\g^{2} + {\gp}^{2}) - 6 \lambda \Big)L_b - 3 \gY^{2} L_f \bigg]
    \nn \\ &
    + T^2 \bigg[ \frac{167}{96}\g^{4} + \frac{1}{288}{\gp}^{4} - \frac{3}{16}\g^{2} {\gp}^{2} + \frac{1}{4}\lambda (3\g^{2}+{\gp}^{2})
    \nn \\ &
    + L_b \Big( \frac{17}{16}\g^{4} - \frac{5}{48}{\gp}^{4} - \frac{3}{16}\g^{2}{\gp}^{2} + \frac{3}{4}\lambda(3\g^{2}+{\gp}^{2}) - 6 \lambda^2 \Big)
    \nn \\ &
    + \Big( c + \ln\left(\frac{3T}{\bmu_{3}}\right) \Big)\Big(  \frac{81}{16}\g^{4} +  3\lambda ( 3\g^{2} + {\gp}^{2})  - 12\lambda^2
    \nn \\ &
    -\frac{7}{16} {\gp}^{4} - \frac{15}{8}\g^{2} {\gp}^{2} \Big)
    \nn \\ &
    - \gY^{2} \Big(\frac{3}{16}\g^{2} + \frac{11}{48}{\gp}^{2} + 2 \gs^{2} \Big)
    + \Big(\frac{1}{12}\g^{4} + \frac{5}{108}{\gp}^{4}\Big)\Nf
    \nn \\ &
    + L_f \Big( \gY^{2} \Big(\frac{9}{16}\g^{2} + \frac{17}{48}{\gp}^{2} + 2 \gs^{2} - 3 \lambda \Big) +\frac{3}{8}\gY^{4}
    \nn\\ &
    - \Big(\frac{1}{4}\g^{4} + \frac{5}{36}{\gp}^{4}\Big) \Nf \Big)
    \nn \\ &
    + \ln(2) \Big( \gY^{2} \Big(-\frac{21}{8}\g^{2} - \frac{47}{72}{\gp}^{2} + \frac{8}{3} \gs^{2} + 9 \lambda \Big) -\frac{3}{2}\gY^{4}
    \nn\\ &
    + \Big(\frac{3}{2}\g^{4} + \frac{5}{6}{\gp}^{4}\Big) \Nf \Big) \bigg] \bigg\}+\Delta \mu_{h,3}^2
    \;.
\end{align}
The $\mu_{h,3}^2$ represents the contribution of the chemical potential to $\mu_{h,3}^2$.
After defining the function
{\setlength{\abovedisplayskip}{1pt}
\setlength{\belowdisplayskip}{1pt}
\begin{equation}
\mathcal{B}(\mu)=\zeta^\prime\left(-1,\frac{1}{2}+\frac{i\mu}{2\pi T}\right)+\zeta^\prime\left(-1,\frac{1}{2}-\frac{i\mu}{2\pi T}\right)-2\zeta^\prime\left(-1,\frac{1}{2}\right),
\end{equation}}
the  $\mu_{h,3}^2$ becomes
\begin{equation}
\begin{aligned}
\Delta \mu_{h,3}^2=&\frac{g_Y^2(\overline{\mu})}{12\pi^2}\mu_B^2-\frac{3g_Y^2}{16\pi^2}\left(\mu_h^2-\frac{\lambda T^2}{2}-\frac{3g^2 T^2}{16}-\frac{g_Y^2 T^2}{4}\right)\mathcal{A}\left(\frac{\mu_B}{3}\right)+\frac{g_Y^2 \mu_B^2}{64\pi^4}\left\{\frac{3}{4}g^2 L_b(\overline{\mu})\right.
\\&\left.-g_Y^2\left[L_f(\overline{\mu})-\mathcal{A}\left(\frac{\mu_B}{3}\right)\right]\right\}-\left[\left(9g_Y^4+\frac{9}{2}g_Y^2 g^2 +16g_Y^2g_s^2\right)\mathcal{A}\left(\frac{\mu_B}{3}\right)\right.\\
&\left.-\left(9g_Y^2+\frac{9}{4}g_Y^2g^2-18\lambda g_Y^2-16g_Y^2g_s^2-\frac{27}{4}g^4\right)16\mathcal{B}\left(\frac{\mu_B}{3}\right)\right]\frac{T^2}{128\pi^2}\\
&+\left(9g_Y^4+\frac{9}{4}g_Y^2 g^2-18\lambda g_Y^2-16g_Y^2g_s^2-\frac{27}{4}g^4\right)\frac{i\mu_B T}{48\pi^3}\ln{\left(\frac{\Gamma\left(\frac{1}{2}-\frac{i\mu_B}{6\pi T}\right)}{\Gamma\left(\frac{1}{2}+\frac{i\mu_B}{6\pi T}\right)}\right)}\\
&+\left[9g_Y^2g^2-6(3g_Y^4-8g_Y^2g_s^2)L_b(\overline{\mu})+9g_Y^4L_f(\overline{\mu})+\left(\frac{9}{2}g_Y^2 g^2+16g_Y^2g_s^2\right)(4\ln{2}-1)\right.\\
&+\left(9g_Y^4+\frac{9}{4}g_Y^2g^2-18\lambda g_Y^2-16g_Y^2 g_s^2-\frac{27}{4}g^4\right)4\gammaE-\left(9g_Y^4+\frac{9}{2}g_Y^2 g^2+16g_Y^2 g_s^2\right)\\
&\left.\mathcal{A}\left(\frac{\mu_B}{3}\right)\right]\frac{\mu_B^2}{1152\pi^4}-\frac{3}{4}g^4\sum_{i=1}^3\left[\frac{T^2}{8\pi^2}\mathcal{B}(\mu_{L_i})+\frac{i\mu_{L_i}T}{16\pi^3}\ln{\left(\frac{\Gamma\left(\frac{1}{2}-\frac{i\mu_{L_i}}{2\pi T}\right)}{\Gamma\left(\frac{1}{2}+\frac{i\mu_{L_i}}{2\pi T}\right)}\right)}+\frac{\mu_{L_i}^2}{32\pi^4}\gammaE\right]
\end{aligned}
\end{equation}

These relations show that the tree level part runs terms of 4d renormalization scale $\overline{\mu}$. After applying the corresponding $\beta-$functions, this running is canceled out by the logarithmic term in the $L_f$ and $L_b$ terms. This leads to a large difference in scale dependence between the tree level and the loop level. However, the scale dependence is not completely removed, but is present in the Debye mass $m_D^2,m_D^{\prime 2}$ and $m_D^{\prime\prime 2}$ at $\mathcal{O}(g^4)$. This scale dependence can be removed by the 2-loop Debye mass at a higher order than $\mathcal{O}(g^4)$ but is beyond the scope of this article.

The chemical potential also includes new terms for effective theories. These terms are read
\begin{align}
\kappa_0=&-\frac{i\pi}{3}g^\prime T^{\frac{5}{2}} \left\{\left(1-\frac{9g^2}{64\pi^2}\right)\sum_{i=1}^3\frac{\mu_{L_i}}{\pi T}\left[1+\left(\frac{\mu_{L_i}}{\pi T}\right)^2\right]-\right.\nn\\
&\left.\left(1-\frac{5g_Y^2}{32\pi^2}-\frac{9g^2}{64\pi^2}-\frac{g_s^2}{2\pi^2}\right)\frac{\mu_B}{\pi T}\left[1+\frac{1}{9}\left(\frac{\mu_B}{\pi T}\right)^2\right]\right\}\;,\\
\rho=&\frac{i}{8\pi}g g_Y^2 T^{3/2}\frac{\mu_B}{\pi T}\;,\\
\rho^\prime=&-\frac{5i}{24\pi} g^\prime g_Y^2 T^{3/2}\frac{\mu_B}{\pi T}\;, \\
\rho_G=&-\frac{i}{8\pi}g^\prime g^2 T^{3/2}\left(\frac{\mu_B}{\pi T}-\sum_{i=1}^3\frac{\mu_{L_i}}{\pi T}\right)\;,
\end{align}
\begin{align}
\alpha_0=&\frac{g^2}{32\pi^2}(N_f \mu_B+\sum_{i=1}^3\mu_{L_i})=0\;,\\
\alpha^\prime=&-\frac{g^{\prime 2}}{32\pi^2}(N_f \mu_B+\sum_{i=1}^3\mu_{L_i})=0\;.
\end{align}
Here, the Chern-Simons terms of $\alpha_0$ and $\alpha^\prime$ are removed due to the chemical potential arrangement.

Finally, when the heavy temporal scalar scalars are integrated out in the second step of dimensional reduction.
The parameters of this light EFT read
\begin{align}\label{lightparameters}
\bar{g}^2_3 =& \g_{3}^{2} \Big( 1 - \frac{\g_{3}^{2}}{6 (4\pi) \mD} \Big)
\;, \\
\bar{g}'^2_3 =& {\gp_{3}}^2
\;, \\
\bar{\mu}^2_{h,3} =& \mu^2_{h,3}
    - \frac{1}{4\pi}\Big(3 h_{1}\mD +  h_{2}\mD' + 8 h_{4}\mD'' \Big)
    + \frac{1}{(4\pi)^2} \bigg( 3\g_{3}^{2}h_{1} - 3 h_{1}^2 - h_{2}^2 - \frac{3}{2} h_{3}^{2}
    \nn \\ &
    + \Big(-\frac{3}{4}\g_{3}^{4} + 12\g_{3}^{2}h_{1} \Big) \ln\Big(\frac{\bmu_{3}}{2\mD} \Big)
    - 6 h_{1}^{2} \ln\Big(\frac{\bmu_{3}}{2\mD} \Big)
    - 2 h_{2}^{2} \ln\Big(\frac{\bmu_{3}}{2\mD'} \Big)
    \nn \\&
    - 3 h_{3}^{2} \ln\Big(\frac{\bmu_{3}}{\mD+\mD'} \Big)
    \bigg)+\Delta \overline{\mu}_h^2
\;, \\
\bar{\lambda}_{3} =& \lambda_{3} - \frac{1}{2(4\pi)}\Big(
      \frac{3 h_{1}^{2}}{\mD}
    + \frac{h_{2}^{2}}{\mD'}
    + \frac{h_{3}^{2}}{\mD+\mD'}
    \Big)+\Delta \overline{\lambda}
    \;,\\
\bar{c}_{6,3} =&
    c_{6,3} + \frac{1}{2(4\pi)}\frac{h_{1}^{3}}{\mD^3}
\;,
\end{align}
where
{\setlength{\abovedisplayskip}{5pt}
\setlength{\belowdisplayskip}{5pt}
\begin{align}
\Delta \overline{\mu}_h^2&=\frac{h_3\kappa_0^2}{m_D^{\prime 4}}-\frac{\rho^\prime \kappa_0}{m_D^{\prime 2}}-\frac{1}{4\pi}\left[(\frac{3g_3^2 g_3^{\prime 2}}{4 m_D}+\frac{4h_2^2}{m_D^\prime})\frac{\kappa_0^2}{m_D^{\prime 4}}-(\frac{3g_3g_3^\prime\rho}{m_D}+\frac{4h_2\rho^\prime}{m_D^\prime})\frac{\kappa_0}{m_D^{\prime 2}}+\frac{3\rho^2}{m_D}+\frac{\rho^{\prime 2}}{m_D^\prime}\right]\\
\Delta \overline{\lambda}&=-\frac{1}{2}\left[(\frac{g_3^2 g_3^{\prime 2}}{4m_D^2}+\frac{4h_2^{\prime 2}}{m_D^{\prime 2}})\frac{\kappa_0^2}{m_D^{\prime 4}}-(\frac{g_3g_3^\prime \rho}{m_D^2}+\frac{4h_2 \rho^\prime}{m_D^{\prime 2}})\frac{\kappa_0}{m_D^{\prime 2
}}+\frac{\rho^2}{m_D^2}+\frac{\rho^{\prime 2}}{m_D^{\prime 2}}\right]
\end{align}}

The light theory depends on an additional renormalization scale $\overline{\mu}_3$ at Eq.\eqref{lightparameters}. This scale is unphysical as well as $\overline{\mu}$. We do not consider the dependence of $\overline{\mu}_3$ since $\overline{\mu}_3$ has a very insignificant effect on the results, and take $\overline{\mu}_3=g^2 T$ is this paper\cite{Croon:2020cgk,Niemi:2021qvp}. 

\begin{figure}[!htp]
  \centering
  \includegraphics[width=0.4\textwidth]{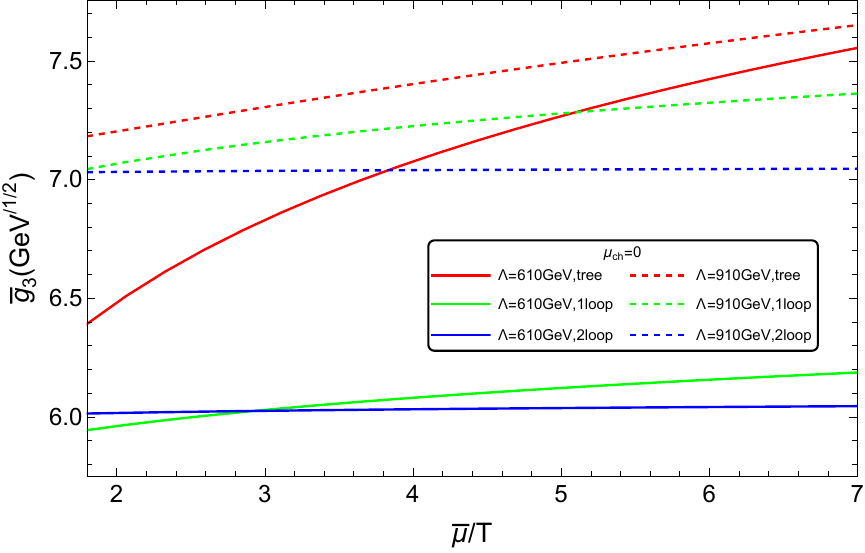}
  \includegraphics[width=0.4\textwidth]{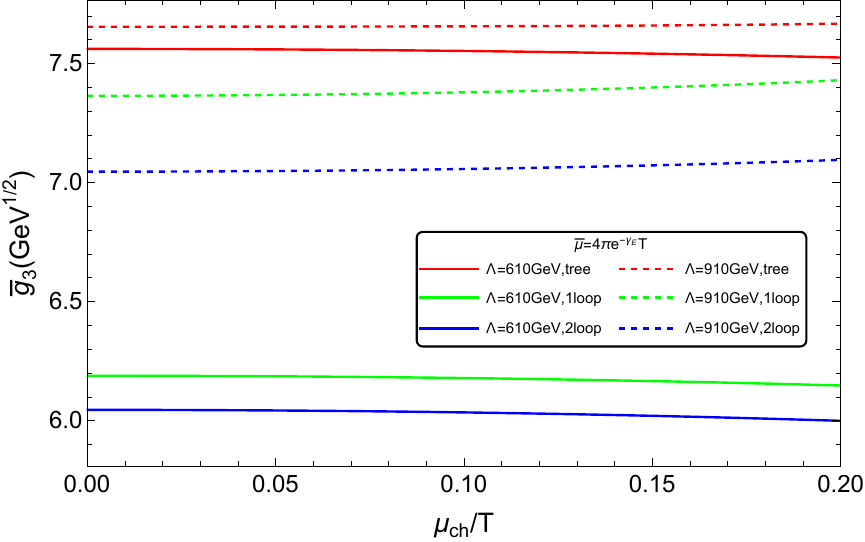}
  \includegraphics[width=0.4\textwidth]{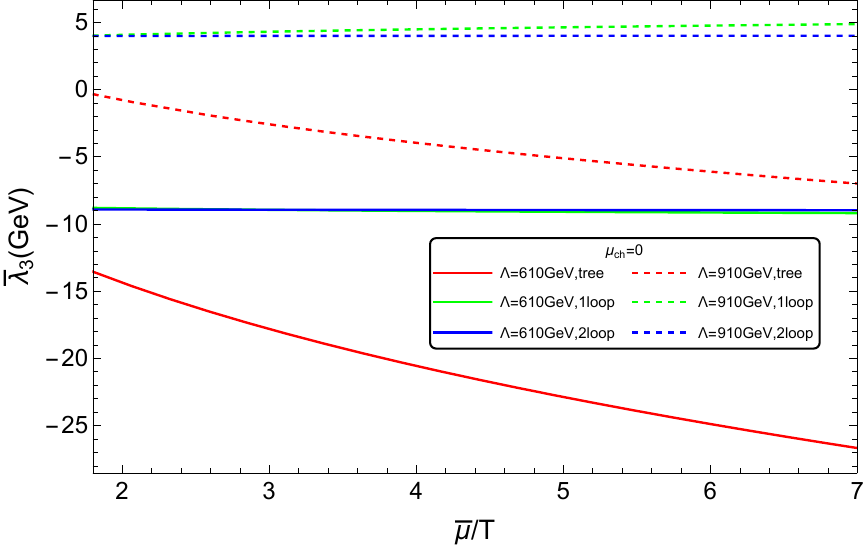}
  \includegraphics[width=0.4\textwidth]{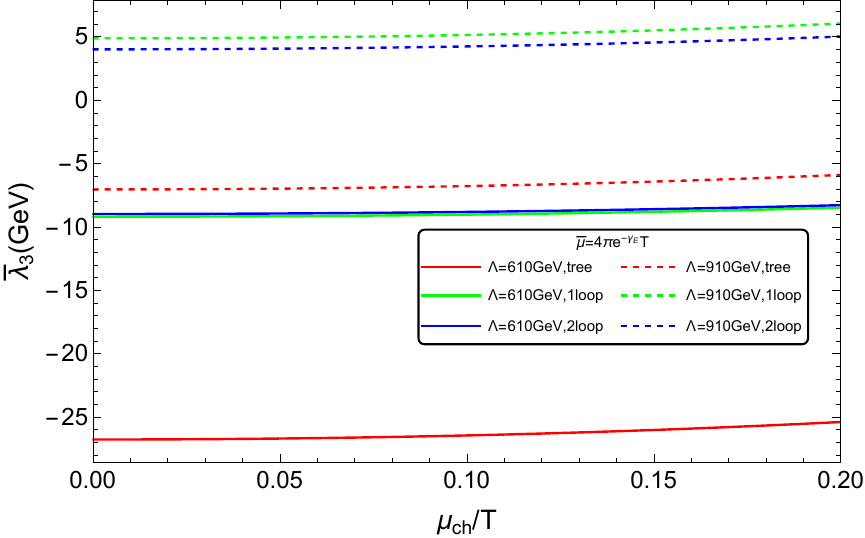}
  \includegraphics[width=0.4\textwidth]{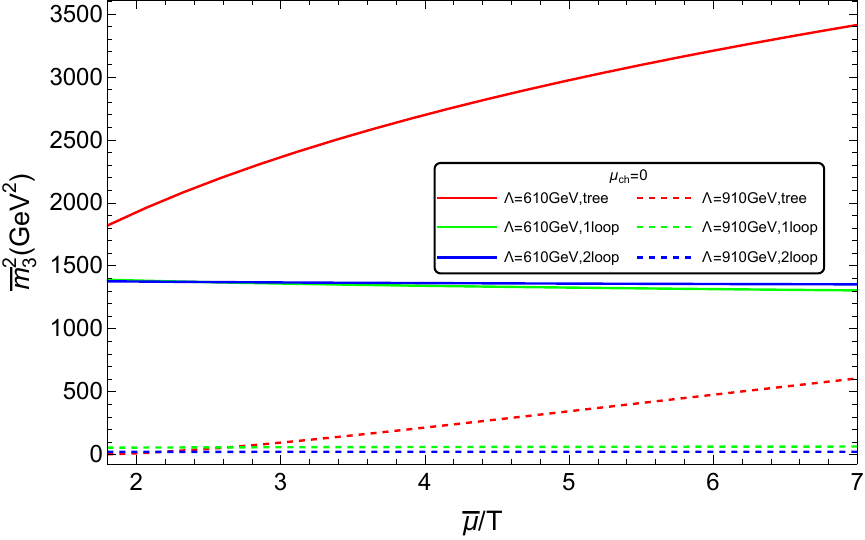}
  \includegraphics[width=0.4\textwidth]{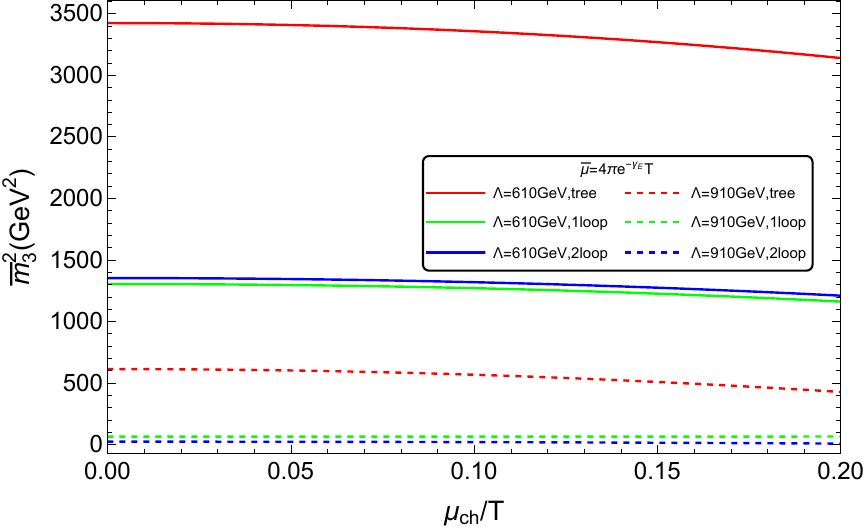}
  \caption{The renormalization scale dependence and effect of chemical potential of $\overline{g}_3,\overline{\lambda}_3$ and $\overline{m}_3^2$ at $T_c$. Left: The renormalization scale dependence of $\overline{g}_3,\overline{\lambda}_3$ and $\overline{m}_3^2$. Right: The effect of chemical potential on $\overline{g}_3,\overline{\lambda}_3$ and $\overline{m}_3^2$ with $\overline{\mu}=4\pi e^{-\gamma-E}T$.}\label{figm3b}
\end{figure}

Fig.\ref{figm3b} shows the renormalization scale dependence and the chemical potential effect on the parameters $\overline{g}_3,\overline{\lambda}_3$ and $\overline{m}_3^2$ at temperature $T_c$. The parameter $\overline{m}_2^3$ at large $\Lambda$ is less than it  at small $\Lambda$, but the opposite is true for $\overline{g}_3^2$ and $\overline{\lambda}_3$. The renormalization scale dependence of these parameters at loop levels is weaker than it is at the tree level. The parameter $\overline{g}_3^2$ is more susceptible to renormalization scale at 1-loop level relative to $\overline{\lambda}_3$ and $\overline{m}_3^2$. These parameters are not sensitive to the chemical potential.

\section{Effective potential at the two-loop level}\label{respotential}

With the definitions of the master integrations of $\mathcal{D}$ and $I_1^3$ being given in Ref.~\cite{Niemi:2020hto}, different part contributions to the effective potential $V_{eff}$ at 2-loop level are:
\begin{align}
{\rm (SSS)} =&
    \frac{3}{4} \phi^2_3 \Big(2 \bar{\lambda}_3 + 3 \bar{c}_{6,3} \phi_3^2 \Big)^2 \mc{D}_{\rmii{SSS}}(m_{\phi,3},m_{\chi,3},m_{\chi,3})
    \nn\\&+ \frac{3}{4} \phi^2_3 \Big( 2 \bar{\lambda}_3 + 5 \bar{c}_{6,3} \phi_3^2 \Big)^2 \mc{D}_{\rmii{SSS}}(m_{\phi,3},m_{\phi,3},m_{\phi,3})
    \;,\\
{\rm (VSS)} =&
    \frac{1}{4} \bar{g}_3^2 \mc{D}_{\rmii{VSS}}(m_{\chi,3},m_{\chi,3},m_{W,3})
    + \frac{1}{4} \bar{g}_3^2 \mc{D}_{\rmii{VSS}}(m_{\phi,3},m_{\chi,3},m_{W,3})
    \nn \\ &+ \frac{1}{8} (\bar{g}_3^2 + \bar{g}_3^{\prime 2})  \mc{D}_{\rmii{VSS}}(m_{\phi,3},m_{\chi,3},m_{Z,3})
     \nn \\ &+ \frac{1}{8} \frac{(\bar{g}_3^2 - \bar{g}_3^{\prime 2})^2}{\bar{g}_3^2 + \bar{g}_3^{\prime 2}} \mc{D}_{\rmii{VSS}}(m_{\chi,3},m_{\chi,3},m_{Z,3})+ \frac{1}{2} \frac{\bar{g}_3^2 \bar{g}_3^{\prime 2}}{\bar{g}_3^2 + \bar{g}_3^{\prime 2}} \mc{D}_{\rmii{VSS}}(m_{\chi,3},m_{\chi,3},0)
    \;,
\end{align}
\begin{align}
{\rm (VVS)} =&
    \frac{1}{8} \bar{g}_3^4 \phi_3^2  \mc{D}_{\rmii{VVS}}(m_{\phi,3},m_{W,3},m_{W,3})+ \frac{1}{16} (\bar{g}_3^2 + \bar{g}_3^{\prime 2})^2 \phi_3^2  \mc{D}_{\rmii{VVS}}(m_{\phi,3},m_{Z,3},m_{Z,3})
    \nn \\ & + \frac{1}{4} \frac{\bar{g}_3^4 \bar{g}_3^{\prime 2} \phi_3^2}{\bar{g}_3^2 + \bar{g}_3^{\prime 2}} \mc{D}_{\rmii{VVS}}(m_{\chi,3},m_W,0)+ \frac{1}{4} \frac{\bar{g}_3^2 \bar{g}_3^{\prime 4} \phi_3^2}{\bar{g}_3^2 + \bar{g}_3^{\prime 2}} \mc{D}_{\rmii{VVS}}(m_{\chi,3},m_{W,3},m_{Z,3})
    \;,\\
{\rm (VVV)} =&
    -\frac{1}{2} \frac{\bar{g}_3^4}{\bar{g}_3^2+\bar{g}_3^{\prime 2}} \mc{D}_{\rmii{VVV}}(m_{W,3},m_{W,3},m_{Z,3})-\frac{1}{2} \frac{\bar{g}_3^2 \bar{g}_3^{\prime 2}}{\bar{g}_3^2+\bar{g}_3^{\prime 2}} \mc{D}_{\rmii{VVV}}(m_{W,3},m_{W,3},0)
    \;,\\
{\rm (VGG)} =&
    -2 \bar{g}_3^2 \mc{D}_{\rmii{VGG}}(m_{W,3})
    -\frac{\bar{g}_3^4}{\bar{g}_3^2+\bar{g}_3^{\prime 2}} \mc{D}_{\rmii{VGG}}(m_{Z,3})
    \;,\\
{\rm (SS)} =&
    -\frac{15}{8} \Big(2\bar{\lambda}_3 + 3 \bar{c}_{6,3} \phi_3^2 \Big) \Big( I^3_1(m_{\chi,3}) \Big)^2- \frac{3}{4} \Big( 9 \bar{c}_{6,3} \phi_3^2 +  2 \bar{\lambda}_3 \Big) I^3_1(m_{\chi,3}) I^3_1(m_{\phi,3})  \nn \\ &
    - \frac{3}{8} \Big( 2\bar{\lambda}_3 + 15 \bar{c}_{6,3} \phi_3^2 \Big) \Big( I^3_1(m_{\phi,3}) \Big)^2
    \;,\\
{\rm (VS)} =&
    - \frac{3}{4} (d-1) \bar{g}_3^2  I^3_1(m_{\chi,3}) I^3_1(m_{W,3})- \frac{1}{4} (d-1) \frac{(\bar{g}_3^2-\bar{g}_3^{\prime 2})^2}{\bar{g}_3^2+\bar{g}_3^{\prime 2}}  I^3_1(m_{\chi,3}) I^3_1(m_{Z,3})
    \\ &
    - \frac{1}{4} (d-1) \bar{g}_3^2  I^3_1(m_{\phi,3}) I^3_1(m_{W,3})- \frac{1}{8} (d-1) (\bar{g}_3^2 + \bar{g}_3^{\prime 2}) I^3_1(m_{\chi,3}) I^3_1(m_{Z,3})
    \nn \\ &
    - \frac{1}{8} (d-1) (\bar{g}_3^2 + \bar{g}_3^{\prime 2})  I^3_1(m_{\phi,3}) I^3_1(m_{Z,3})
    \;,\\
{\rm (VV)} =&
    - \frac{1}{2} \bar{g}_3^2 \mc{D}_{\rmii{VV}}(m_{W,3}, m_{W,3}) - \frac{\bar{g}_3^4}{\bar{g}_3^2+\bar{g}_3^{\prime 2}}\mc{D}_{\rmii{VV}}(m_{W,3}, m_{Z,3}) \;.
\end{align}

\begin{figure}[!htp]
    \centering
    \includegraphics[width=0.4\textwidth]{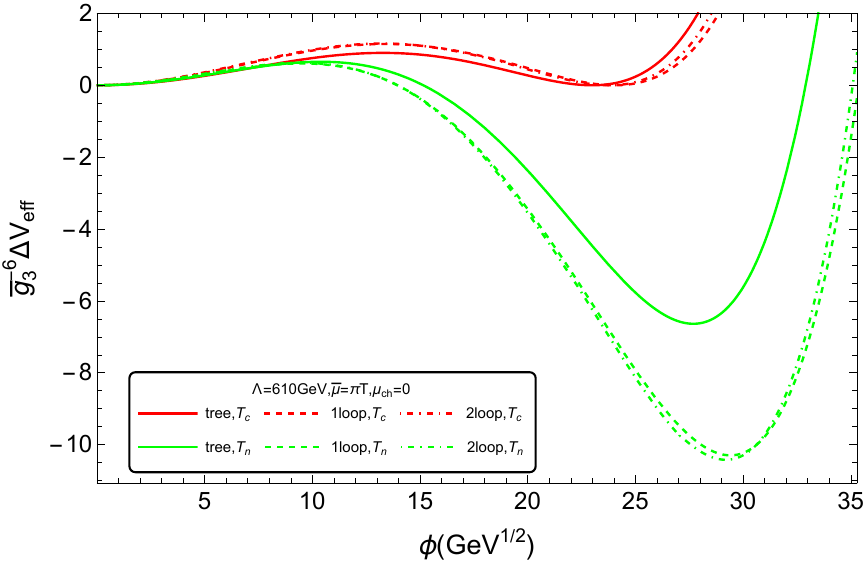}
    \includegraphics[width=0.4\textwidth]{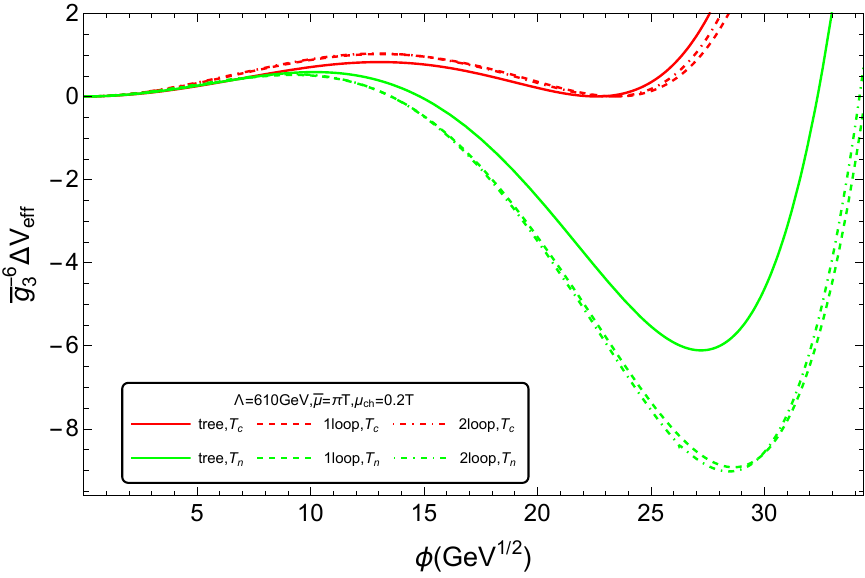}
    \includegraphics[width=0.4\textwidth]{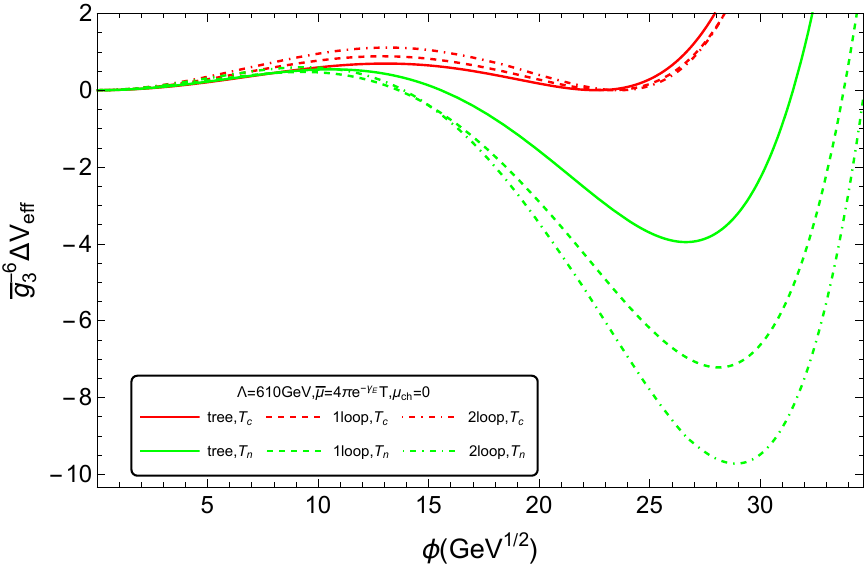}
    \includegraphics[width=0.4\textwidth]{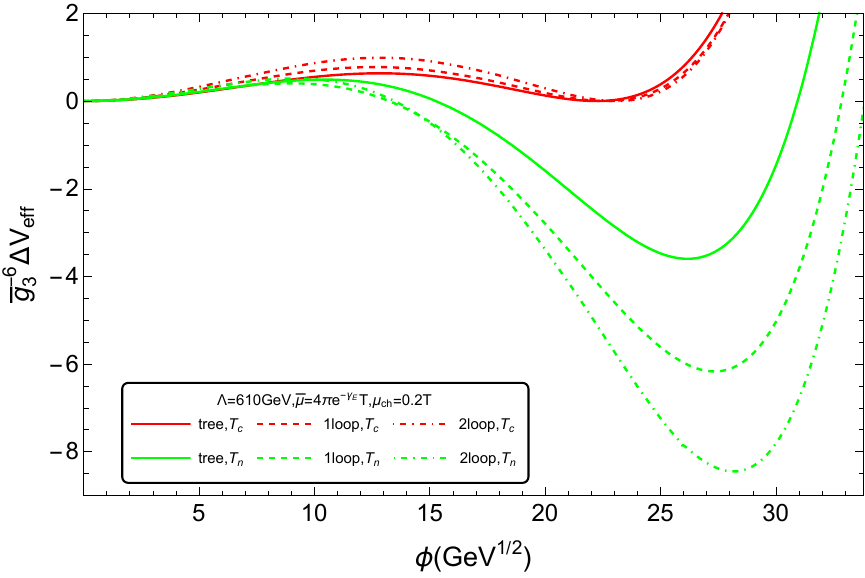}
    \caption{The effective potential at $\Lambda=610$ GeV with different renormalization scales and chemical potentials.}
    \label{figpotential610}
\end{figure}

\begin{figure}[!htp]
    \centering
    \includegraphics[width=0.4\textwidth]{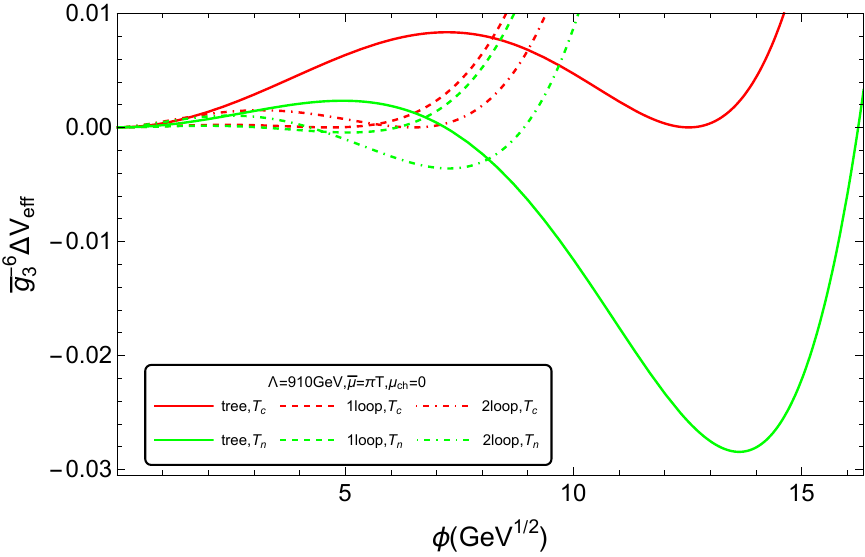}
    \includegraphics[width=0.4\textwidth]{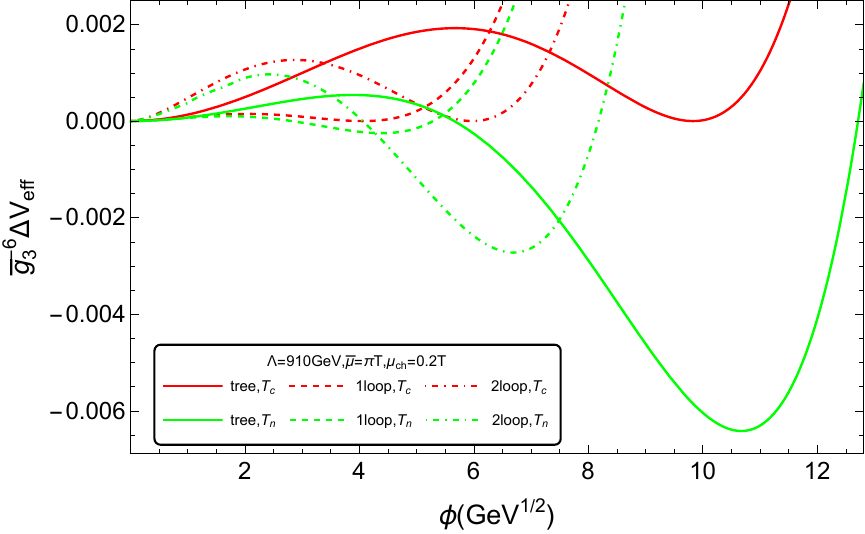}
    \includegraphics[width=0.4\textwidth]{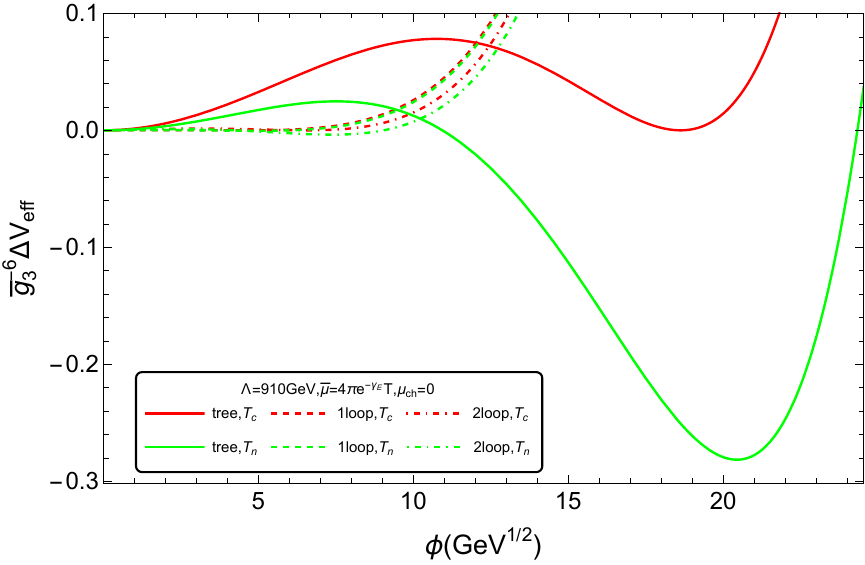}
    \includegraphics[width=0.4\textwidth]{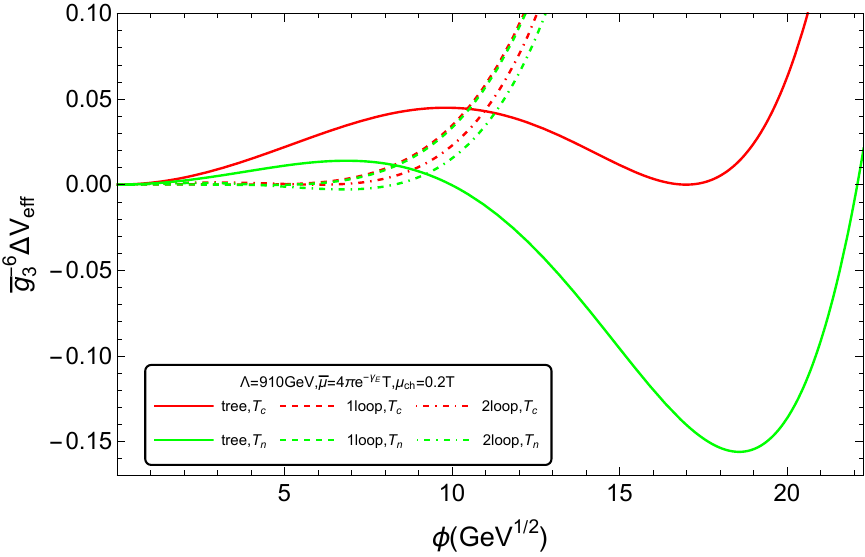}
    \caption{The effective potential at  $\Lambda=910$ GeV with different renormalization scales and chemical potentials.}
    \label{figpotential910}
\end{figure}

The Figs.(\ref{figpotential610}, \ref{figpotential910}) show the effective potential at $T_c$(red) and $T_n$(green) with $\Lambda=610$ GeV and $910$ GeV.
Where, we can find that the renormalization scale dependence of the thermal effective potential decreases from tree-level to two-loop level, and the chemical potential can reduce the magnitude of the potential barrier $\bar{g}^{-6}_3\Delta V_{eff}$.
For the case of $\Lambda=610$ GeV, we can see that the magnitude of $|\Delta V_{eff}|$ at loop levels is larger than that at tree level at the nucleation temperature $T_n$, and the $|\Delta V_{eff}|$ at loop levels is smaller than it at tree level with $\Lambda=910$GeV. This means that, in the range of $\Lambda=590-910$GeV, there exists a value of $\Lambda$ where the loop levels contribution cancels that from the tree level, i.e.,
$|\Delta V_{eff}^{tree}|=|\Delta (V_{eff}^{1loop}+V_{eff}^{2loop})|$.

\section{Sphaleron energy}\label{resspe}

The sphaleron energy has the form
\begin{equation}\label{spe1}
E_{sph}(T)=\int d^3 x \left[-\frac{1}{4}G_{ij}^a G_{ij}^a-\frac{1}{4}F_{ij}F_{ij}+(D_i \Phi)^\dag (D_i \Phi)+V_{eff}\right]
\end{equation}
and  $G_{ij},F_{ij},D_i$ are defined in Eq.\eqref{action3dheavy}.
Then, we consider sphaleron energy with $U(1)$ gauge field. For $\theta_w\neq 0$, the $U(1)$ gauge field will contribute to sphaleron energy, and the spherical symmetry reduces to an axial symmetry\cite{Comelli:1999gt,DeSimone:2011ek,Manton:1983nd,Klinkhamer:1990fi}. The $SU(2)$ matrix is defined as
\begin{equation}\label{su2matrix}
 U^{\infty}(\nu,\theta,\phi) =\begin{pmatrix}
e^{i\nu}(\cos\nu-i \sin\nu\cos\theta) & e^{i\phi}\sin\nu\sin\theta \\
  -e^{-i\phi}\sin\nu\sin\theta &  e^{-i\nu}(\cos\nu+i \sin\nu\cos\theta),
\end{pmatrix}
\end{equation}
then we can define
\begin{equation}
U^{\infty -1}dU^{\infty}=\sum_{a=1}^{3}F_a\frac{\tau^a}{2i},
\end{equation}
where
\begin{equation}
\begin{aligned}
F_1=&-2 \sin\nu \left[\cos{\theta} \cos^2{\nu} \sin{\phi}-2 \sin ^2{(\theta/2)} \sin{\nu} \cos{\nu} \cos{\phi}+\sin^2{\nu}\sin{\phi}\right]d\theta\\
&-2\sin{\nu}\left[\cos{\phi}\sin{\theta} -\cos{\phi}\sin{\theta}\sin^2{\nu} +\cos{\theta}\cos{\phi}\sin{\theta}\sin^2{\nu}\right.\\
&\left.-2\cos{\nu}\sin^2{(\theta/2)}\sin{\theta}\sin{\nu}\sin{\phi}\right] d\phi ,\\
F_2=&-2\sin{\nu}\left[\cos{\phi}(\cos{\theta}\cos^2{\nu}+\sin^2{\nu}+\sin^2{(\theta/2)}\sin{2\nu}\sin{\phi}) \right]d\theta\\
&+2\sin{\theta}\sin{\nu}\left[\cos{\phi}\sin^2{(\theta/2)}\sin{2\nu}+(1+(-1+\cos{\theta})\sin^2{\nu})\sin{\phi}\right]d\phi ,\\
F_3=&2(\sin\nu\cos\nu\sin\theta d\theta+\sin^2 \nu\sin^2 \theta d\phi ) .\\
\end{aligned}
\end{equation}
After set $\nu=\pi/2$, the function $F_i$ becomes
\begin{equation}
\begin{aligned}
F_1&=-2\sin \phi d\theta -\sin 2\theta \cos \phi d\phi ,\\
F_2&=-2\cos \phi d\theta +\sin 2\theta \sin \phi d\phi ,\\
F_3&=2 \sin^2{\theta}d\phi .\\
\end{aligned}
\end{equation}
The configuration of gauge and Higgs fields is expressed as
\begin{equation}\label{ansatzes2}
\begin{aligned}
g^\prime B_i d x^i&=(1-f_0(\xi))F_3 ,\\
g A_i^a\tau^a d x^i&=(1-f(\xi))(F_1\tau^1+F_2\tau^2)+(1-f_3(\xi))F_3\tau^3 ,\\
\phi&=\frac{v(T)}{\sqrt{2}}\begin{pmatrix}
                             0 \\
                             h(\xi)
                           \end{pmatrix} ,\\
\end{aligned}
\end{equation}

and the sphaleron energy read as
\begin{equation}
\begin{aligned}
E_{sph}^{EW}(\nu,\theta,\phi,T)=&\frac{v}{g}\int \xi^2\sin\theta d\xi d\theta d\phi\left[\sin^2\nu \frac{1}{\xi^2}\{4f^{\prime 2}+2\sin^2\theta(f_3^{\prime 2}-f^{\prime 2})\}\right.\\
&+\sin^4\nu\frac{8}{\xi^4}\left\{\sin^2\theta f_3^2(1-f)^2+\cos^2\theta(f(1-f)+f-f_3)^2\right\}\\
&+\frac{g}{g^\prime}\left\{\sin^2\nu\sin^2\theta\frac{2}{\xi^2}f_0^{\prime^2}+\sin^4\nu\cos^2\theta\frac{8}{\xi^4}(1-f_0)^2\right\}\\
&+\frac{1}{2}h^{\prime 2}+\sin^2\nu\frac{h^2}{2\xi^2}\left\{\sin^2\theta(f_0-f_3)^2+(2-\sin^2\theta)(1-f)^2\right\}\\
&\left.+\frac{\xi^2}{g^2 v(T)^4}V(h,T)\right].
\end{aligned}
\end{equation}

\begin{figure}[!htp]
    \centering
\includegraphics[width=0.8\textwidth]{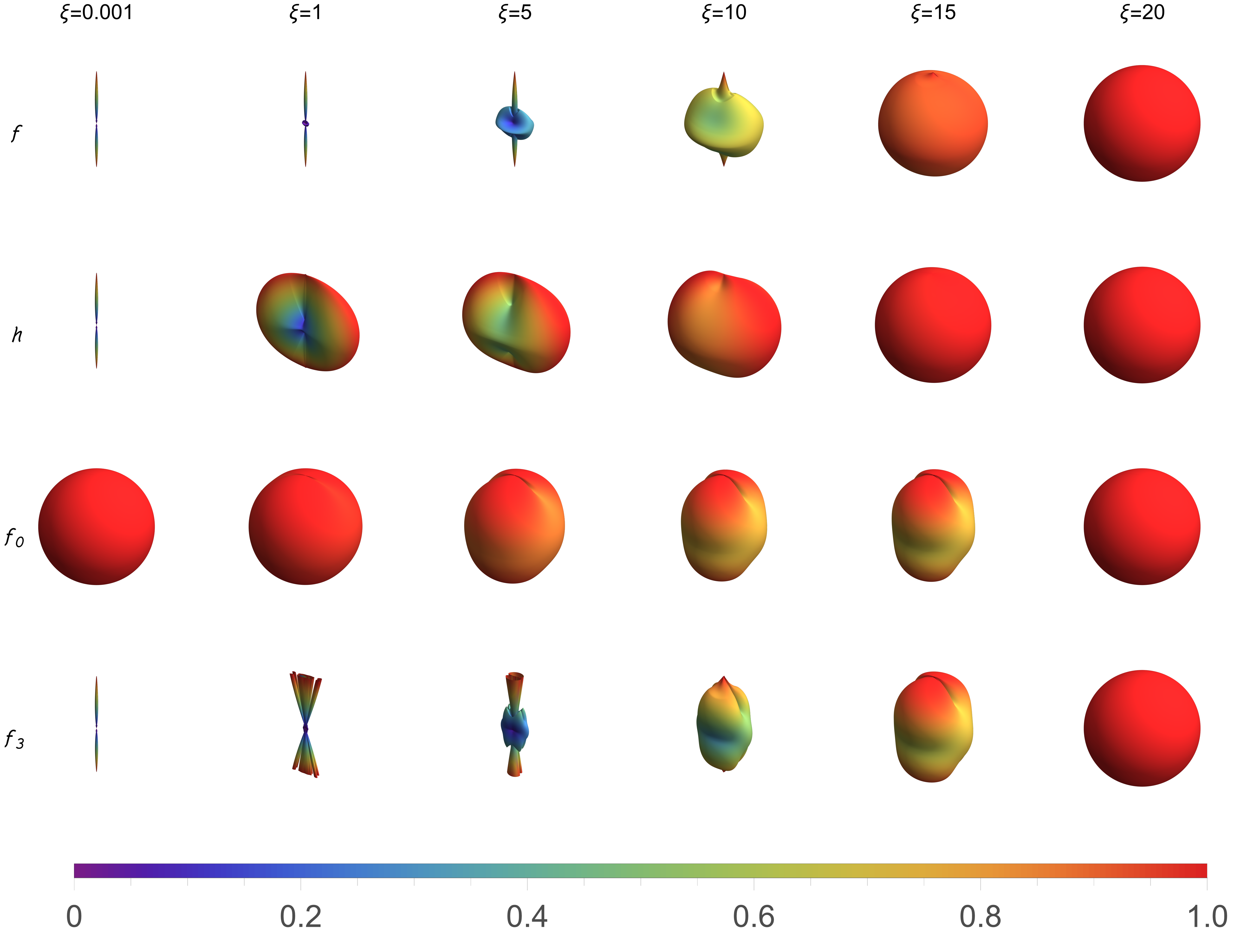}
    \caption{The behaviors of $f,h,f_0,f_3$ as functions of variables $\xi$,$\theta$ and $\nu$ with $\theta\in [0,\pi]$ and $\nu\in [-\pi,\pi]$, here $\Lambda=610$ GeV and we take $\overline{\mu}=4\pi E^{-\gamma_E}T $ and $\mu_{ch}=0$ as an expample. }
    \label{figfh}
\end{figure}

After integrating over the variables $\theta$  and $\phi$, the sphaleron energy become
\begin{equation}\label{spe4}
\begin{aligned}
E_{sph}^{EW}(\nu,T)=&\frac{4\pi v}{g}\frac{v(T)}{v}\int_{0}^{\infty}d\xi \left\{\sin^2{\nu}\left(\frac{8}{3}f^{\prime 2}+\frac{4}{3}f^{\prime 2}_3\right)+\sin^4{\nu}\frac{8}{\xi^2}\left[\frac{2}{3}f_3^2(1-f)^2\right.\right.\\
&\left.+\frac{1}{3}(f(1-f)+f-f_3)^2\right]+\frac{4}{3}\left(\frac{g}{g^\prime}\right)^2\left[\sin^2{\nu}f_0^{\prime 2}+\sin^4{\nu}\frac{2}{\xi^2}(1-f_0)^2\right]\\
&+\frac{1}{2}\xi^2h^{\prime 2}+\sin^2{\nu}h^2\left[\frac{1}{3}(f_0-f_3)^2+\frac{2}{3}(1-f)^2\right]+\left.\frac{\xi^2}{g^2 v(T)^4}V(h,T)\right\}.\\
\end{aligned}
\end{equation}

The function $f(\xi),h(\xi),f_0(\xi)$ and $f_3(\xi)$ in eq.\eqref{spe4} can be obtained by solve equations\eqref{dipolefuns}
\begin{equation}\label{dipolefuns}
\begin{aligned}
&f^{\prime \prime}+\frac{1-f}{4\xi^2}\left[8\sin^2 \nu(f(f-2)+f_3+f_3^2)+\xi^2h^2\right]=0 ,\\
&f^{\prime \prime}_3-\frac{2}{\xi^2}\sin^2 \nu\left[3f_3+f(f-2)(1+2f_3)\right]-\frac{h^2}{4}(f_3-f_0)=0 ,\\
&f^{\prime \prime}_0+\sin^2 \nu\frac{g^{\prime 2}}{4g^2} h^2(f_3-f_0)+2\frac{1-f_0}{\xi^2}=0 ,\\
&h^{\prime \prime}+\frac{2}{\xi}h^\prime-\sin^2 \nu\frac{2}{3\xi^2}h\left[2(f-1)^2+(f_3-f_0)^2\right]-\frac{1}{g^2v(T)^4}\frac{\partial V(h,T)}{\partial h}=0
\end{aligned}
\end{equation}
with boundary condition
\begin{equation}
\begin{aligned}
f(\xi)=0, h(\xi)=0, f_3(\xi)=0, f_0(\xi)=1&,\qquad \xi\rightarrow 0\\
f(\xi)=1, h(\xi)=1, f_3(\xi)=1, f_0(\xi)=1&,\qquad \xi\rightarrow \infty\\
\end{aligned}
\end{equation}

\begin{figure}[!htp]
    \centering
    \includegraphics[width=0.4\textwidth]{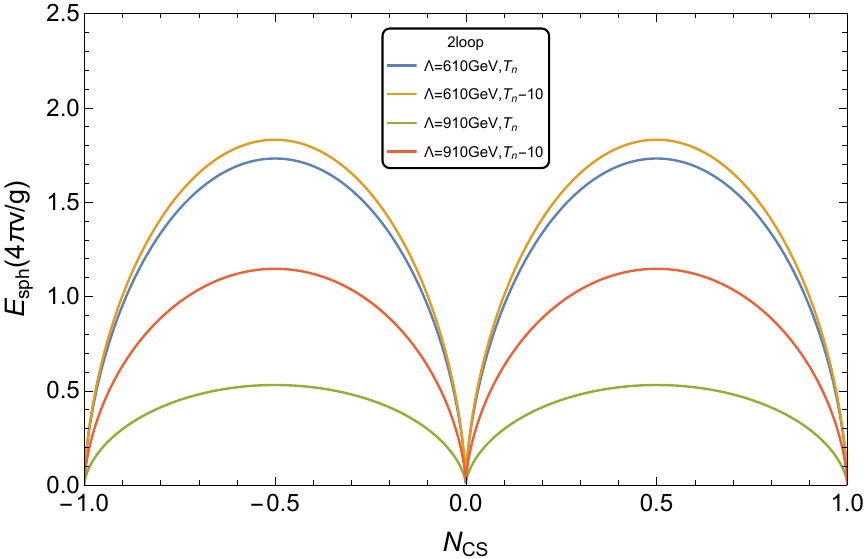}
    \includegraphics[width=0.4\textwidth]{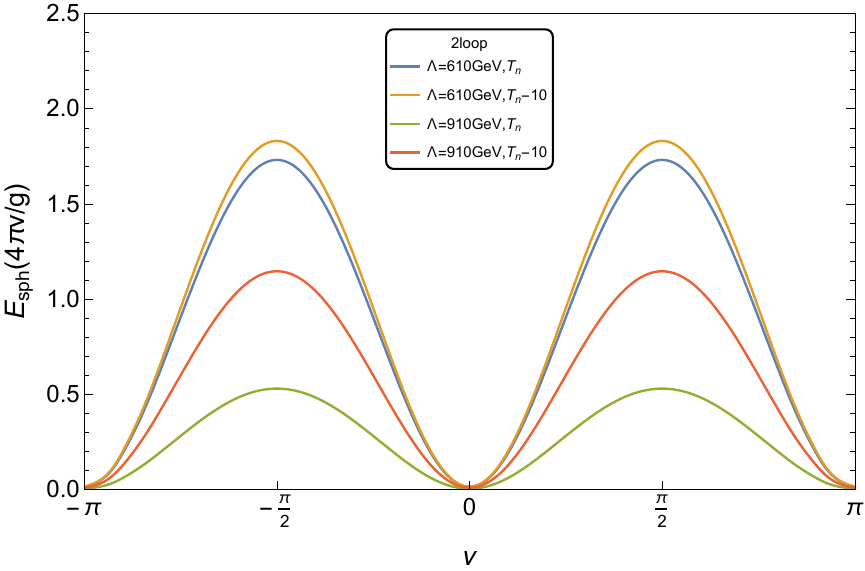}
    \caption{The sphaleron energy as a function of $\nu$ and $N_{CS}$. The left plot is the sphaleron energy as a function of $N_{CS}$ and in the right plot the sphaleron energy is shown as a function of $\nu$. The lowest value is the vanishing sphaleron energy and each highest value in the barrier between two adjacent vacua corresponds to a sphaleron solution. }
    \label{figncs}
\end{figure}

A solutions of Eq.\eqref{dipolefuns} without integrating over $\theta$ are shown in Fig.~\ref{figfh}, one can find that all the configurations of $f,h,f_0,f_3$ finally approach to spherical at large $\xi$.
As shown in Fig.\ref{figncs}, there is a periodic structure of the sphaleron energy with Chern-Simons number(left) and $\nu$(right). The integral value of $N_{CS}$ need $\sin{2\nu}=0(x=n\pi)$, this condition will remove lots of terms of sphaleron energy Eq.\eqref{spe3} and lead a periodic structure to the sphaleron energy with $N_{CS}$ and $\nu$. In this study, we set $\nu=\pi/2$ to maximize the sphaleron energy.

\begin{figure}[!htp]
    \centering   \includegraphics[width=0.4\textwidth]{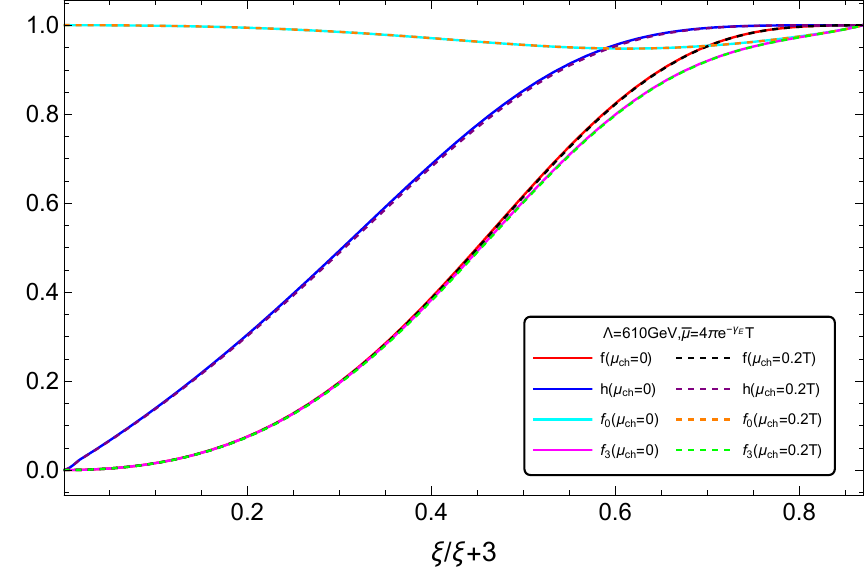}
    \includegraphics[width=0.4\textwidth]{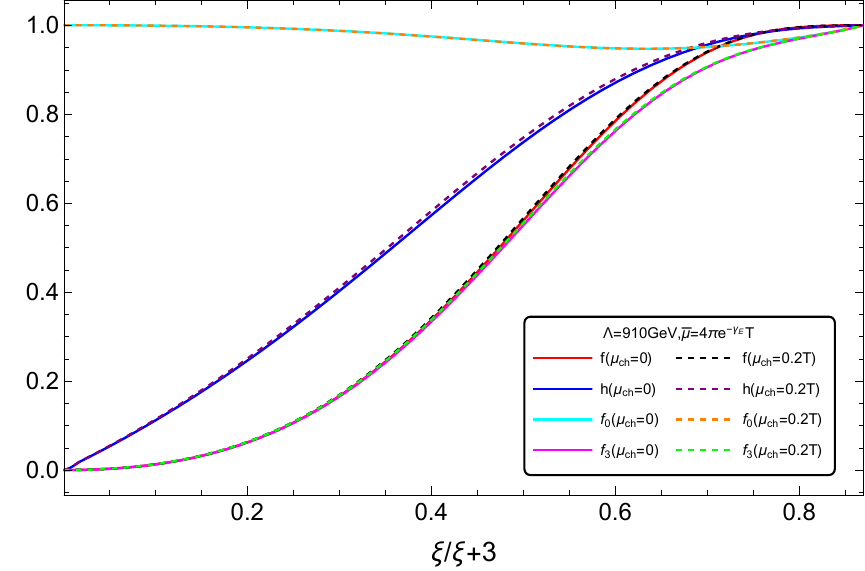}
    \caption{Examples of solutions to eq.\eqref{dipolefuns} at two-loop order with $\nu=\pi/2$ for the cases of $\Lambda=600$ GeV (left plot) and $\Lambda=900$ GeV (right plot).}
    \label{figsolutions}
\end{figure}

\begin{figure}[!htp]
  \centering
\includegraphics[width=0.4\textwidth]{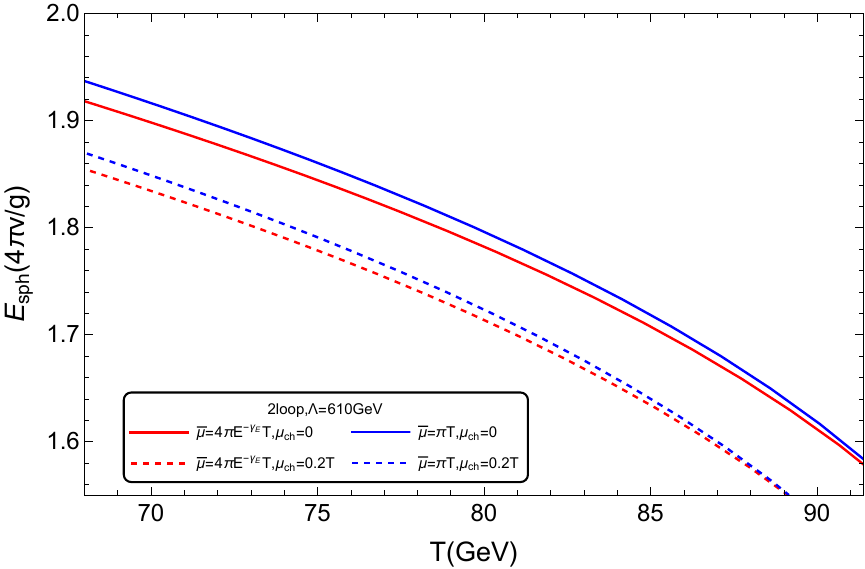}
\includegraphics[width=0.4\textwidth]{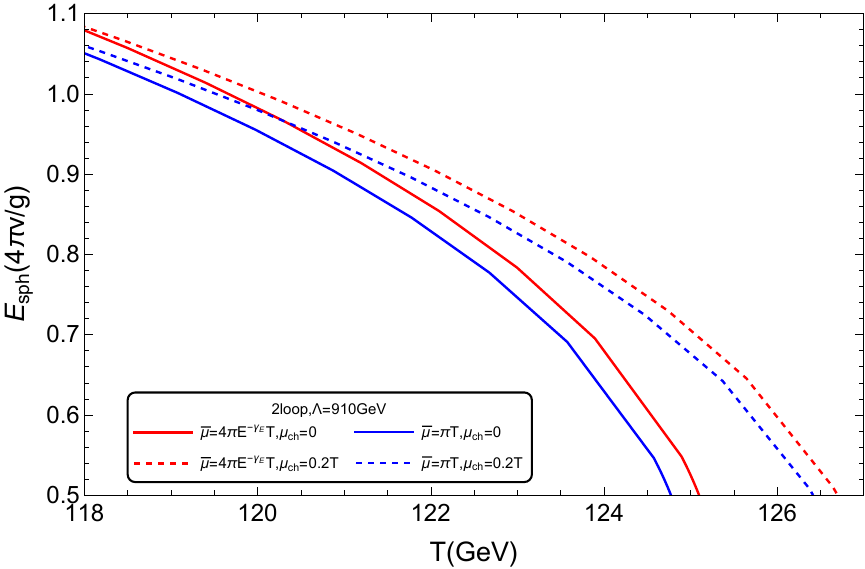}
  \caption{The sphaleron energys as functions of temperature with $\Lambda=610$ GeV (left panel) and $910$ GeV (right panel) at 2-loop level.}\label{figspts}
\end{figure}

Fig.\ref{figsolutions} shows that the renormalization scale and chemical potential have a negligible effect on the solution of Eq.\eqref{dipolefuns} where both angles of $\theta$ and $\phi$ are integrated out. The figure.\ref{figspts} shows that the decrease in sphaleron energy with increasing temperature for two benchmarks of $\Lambda=610$ GeV and $\Lambda=900$ GeV when  $\nu=\pi/2$, that shows that the chemical potentials contribution will be enlarged at high NP scale.

When the $U(1)$ gauge field has contributed to sphaleron energy, a magnetic dipole moment will arise.
The dipole energy with a constant magnetic field $B$ has an additional contribution to sphaleron energy with
\begin{equation}
E_{dip}=-\mu^{(1)}B ,
\end{equation}
where the dipole moment $\mu^{(1)}$  was given by
\begin{equation}\label{dipolexi}
\mu^{(1)}=\frac{2\pi}{3}\frac{g^\prime}{g^3 v}\int_{0}^{\infty}\mathrm{d}\xi \, \xi^2 h^2(\xi)[1-f(\xi)]\;.
\end{equation}
use $x=\xi/(\xi+3)$, it can be rewritten as
\begin{equation}\label{dipolex}
\mu^{(1)}=\frac{2\pi}{3}\frac{g^\prime}{g^3 v}\int_{0}^{1}\mathrm{d}x \, \left(-\frac{3}{-1+x}+\frac{3x}{(-1+x)^2}\right)\left(-\frac{3x}{-1+x}\right)^2 h^2(x)[1-f(x)]\;.
\end{equation}
Then the sphaleron energy with magnetic field becomes: $E_{sph}=E^{EW}_{sph}+E_{dip}$.
Therefore, the magnetic field would threaten the EWBG by changing the electroweak sphaleron rate around the PT~\cite{DeSimone:2011ek}.

\begin{figure}[h]
    \centering
    \includegraphics[width=0.47\textwidth]{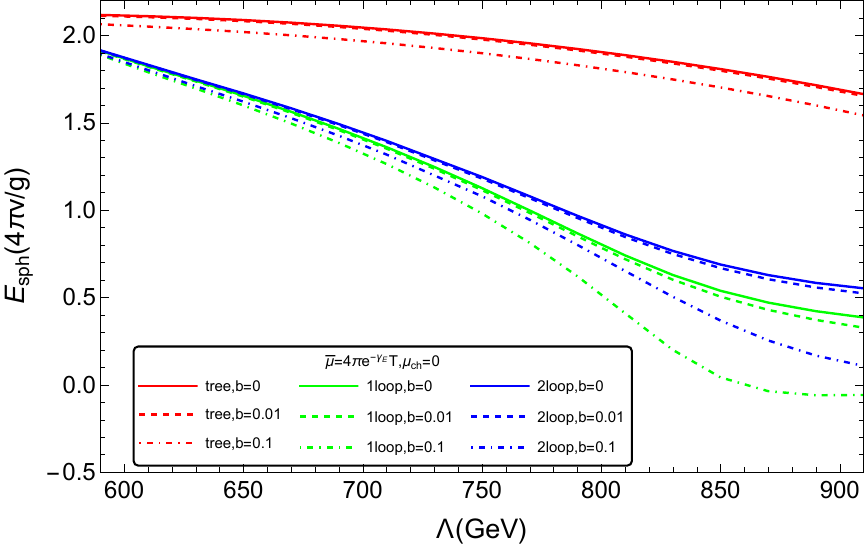}
    \includegraphics[width=0.47\textwidth]{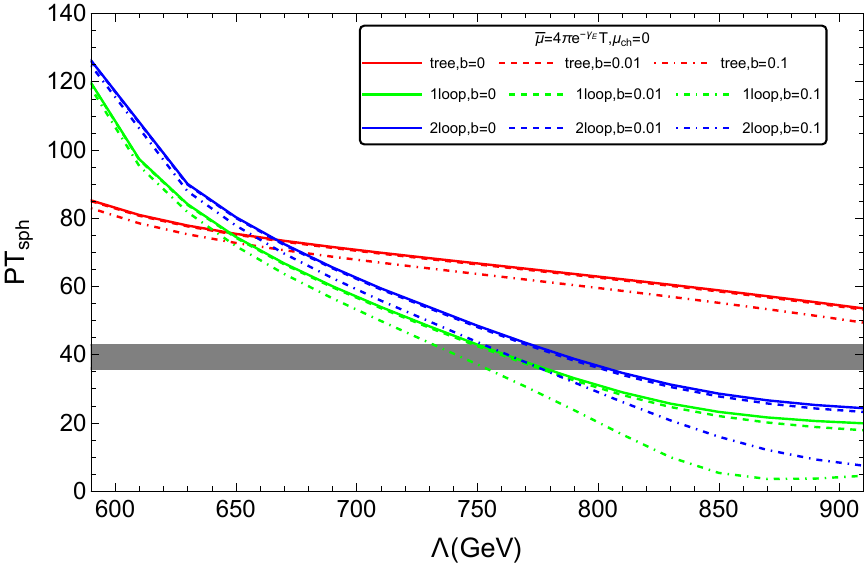}
    \caption{The effect of the dipole moment of $E_{sph}$ (left plot) and $PT_{sph}$ (right plot) with magnetic field $B=b T^2$ at $T_n$. We set the renormalization scale at $\overline{\mu}=4\pi T e^{-\gamma_E}T$ with null chemical potential.}
    \label{figb}
\end{figure}

In Fig.\ref{figb}, we show the effect of dipole moment on $E_{sph}$ and $PT_{sph}$ with the magnetic field being $B=b T_n^2$. We take $b=0.1,0.01$ as an illustration since the magnetic field generated at PT is found to be around the magnitude~\cite{Ahonen:1997wh,Zhang:2019vsb,Stevens:2012zz,Yang:2021uid,Di:2020kbw}. The nucleation temperature $T_n$ is obtained by Fig.\ref{figtn} and we do not consider the effects of the magnetic field on the PT dynamics since it was found not to drive first-order PT~\cite{Kajantie:1998rz}
, while it might yield W condensate see Refs.\cite{Elmfors:1998wz,Chernodub:2022ywg} for example. The effect of the dipole moment on the quantities of $E_{sph}$ and $PT_{sph}$ at small $\Lambda$ region is negligible, but it becomes very significant at large $\Lambda$ region.
The magnetic field can modify the restriction on $\Lambda$ by changing the washout avoidance condition. At the 2-loop level with $\overline{\mu}=4\pi e^{-\gamma_E} T$ and $\mu_{ch}=0$, the magnetic field changes the restriction on $\Lambda$ from $770-800$ GeV ($B=0$) to $750-780$ GeV ($B=0.1T^2$). This result is similar to the scenario of $\mu_{ch}=0.2T$ in Fig.\ref{figpt}, and when $\Lambda\gtrsim 800$ GeV, the effect of magnetic field on the $PT_{sph}$ increases with $\Lambda$ and grows to be larger than that of the chemical potential. This implies that the role of the magnetic field in the weak PT is non-negligible relative to the chemical potential.


\bibliography{reference}

\end{document}